\documentclass[12pt]{article}
\usepackage{amssymb, amsmath,amsthm}
\usepackage[dvips]{graphicx}

\date{}

\begin{document}

\title
{\large{Magnetic Relativistic Schr\"odinger Operators
and \\Imaginary-time Path Integrals }
\thanks{Mathematical Physics, Spectral Theory and Stochastic Analysis, Operator
Theory: Advances and Applications 232, pp. 247--297, Springer/Birkha\"user 2013}
}
\author{{Takashi Ichinose}
\\\normalsize{
Department of Mathematics, Kanazawa University, Kanazawa, 920-1192, Japan} 
\\ \normalsize{E-mail: ichinose@kenroku.kanazawa-u.ac.jp}\\
\footnotesize{\sl }
}

\maketitle

\abstract
{Three magnetic relativistic Schr\"odinger operators corresponding to 
the classical relativistic Hamiltonian symbol with magnetic vector and 
electric scalar potentials are considered, dependent on how to quantize 
the kinetic energy term $\sqrt{(\xi-A(x))^2 +m^2}$.
We discuss their difference in general and their coincidence 
in the case of constant magnetic fields, and also study whether they are 
covariant under gauge transformation. 
Then results are reviewed on 
path integral representations for their respective imaginary-time 
relativistic Schr\"odinger equations, i.e. heat equations, by means of 
the probability path space measure related to the L\'evy process concerned. }



\makeatletter
\renewcommand{\theequation}{%
  \thesection.\arabic{equation}}
\@addtoreset{equation}{section}
\makeatother

\newtheorem{dfn}{Definition}[section]
\newtheorem{thm}[dfn]{Theorem}
\newtheorem{prp}[dfn]{Proposition}
\newtheorem{lemma}[dfn]{Lemma}

\bigskip\noindent
CONTENTS

\noindent
1. Introduction

\noindent
2. Three magnetic relativistic Schr\"odinger operators

2.1. Their definition and difference

2.2. Gauge-covariant or not

\noindent
3. More general definition of magnetic relativistic Schr\"odinger 
   operators and their selfadjointness

3.1. The most general definition of $H_A^{(1)}$, $H_A^{(2)}$ and $H_A^{(3)}$ 

3.2. Selfadjointness with negative scalar potentials 

\noindent
4. Imaginary-time path integrals for magnetic relativistic Schr\"odinger operators

4.1. Feynman--Kac--It\^o type formulas for magnetic relativistic 
Schr\"odinger operators

4.2. Heuristic derivation of path integral formulas

\noindent
5. Summary

\noindent
\quad References

\section{Introduction} 

We consider the quantized operator $H := H_A + V$ corresponding to 
the symbol of the classical relativistic Hamiltonian 
\begin{equation}
\sqrt{(\xi-A(x))^2 +m^2} +V(x)\,, \qquad 
                 (\xi,x) \in {\bf R}^d \times {\bf R}^d\,, 
\end{equation}
for a {\sl relativistic} spinless particle of mass $m$ 
under influence of the {\sl magnetic} vector potential $A(x)$ 
and electric scalar potential $V(x)$ being, respectively,   
an ${\bf R}^d$-valued function and a real-valued function on space ${\bf R}^d$. 
This $H$ is effectively used in the situation where one may ignore 
quantum-field theoretic effects like particles creation and annihilation 
but should take relativistic effect into consideration. 
Throughout, the speed $c$ of light and the constant $\hbar :=h/2\pi$, 
the Planck's constant $h$ divided by $2\pi$, are taken to be equal to 1.

If the vector potential $A(x)$ is absent, i.e. $A(x)$ is a constant vector, 
we can define $H$ as 
$H = H_0 +V$, where $H_0 :=\sqrt{-\Delta+m^2}$ with $-\Delta$ the Laplace 
operator in ${\bf R}^d$ and $V$ is a multiplication operator by 
the function $V(x)$. We can then realize not only these $H_0$ and $V$
but also their sum $H_0 +V$ as selfadjoint operators in $L^2({\bf R}^d)$,  
so long as we consider some class of reasonable scalar potential functions 
$V(x)$.
However, when the vector potential $A(x)$ is present, the definition of $H$
involves some sort of ambiguity. In fact, in the literature there are 
three kinds of quantum relativistic Hamiltonians dependent on how to quantize 
the kinetic energy symbol $\sqrt{(\xi-A(x))^2 +m^2}$ to get the first term 
$H_A$ of $H$, the kinetic energy operator.

In this article, we will treat these three quantized operators 
$H^{(1)} = H^{(1)}_A + V$, $H^{(2)}= H^{(2)}_A + V$ 
and $H^{(3)}= H^{(3)}_A + V$ 
corresponding to the classical relativistic Hamiltonian symbol (1.1) 
which have the following kinetic energy parts 
$H^{(1)}_A,\, H^{(2)}_A$ and $H^{(3)}_A$.
At first here, at least in this introduction, we assume for simplicity that 
$A(x)$ is a {\it smooth} ${\bf R}^d$-valued function which together with
all its derivatives is bounded and that 
$V(x)$ is a real-valued bounded function. 

The first two $H^{(1)}_A$ and $H^{(2)}_A$ are to be defined as 
pseudo-differential operators through oscillatory integrals. 
For a function $f$ in $C_0^{\infty}({\bf R}^d)$ 
put
\begin{eqnarray}
(H^{(1)}_A f)(x) 
\!\!&\!:=\!&\!\! \frac1{(2\pi)^{d}} \int\int_{{\bf R}^d\times {\bf R}^d} 
   e^{i(x-y)\cdot\xi}\sqrt{\Big(\xi-A\big(\frac{x+y}{2}\big)\Big)^2 +m^2}
   \,f(y) dyd\xi, \qquad \\ 
(H^{(2)}_A f)(x) 
\!\!&\!:=\!&\!\! \frac1{(2\pi)^{d}}\! \int\int_{{\bf R}^d\times {\bf R}^d} 
   e^{i(x-y)\cdot \xi} \nonumber\\
&&\qquad \times
  \sqrt{\Big(\xi- \int_0^1 A((1-\theta)x+\theta y)d\theta\Big)^2 +m^2}\,f(y) 
   dyd\xi.    
\end{eqnarray}
The third $H^{(3)}_A$ is defined as the square root of the nonnegative 
selfadjoint operator $(-i\nabla-A(x))^2+m^2$ in $L^2({\bf R}^d)$:
\begin{equation}%
H^{(3)}_A := \sqrt{(-i\nabla-A(x))^2+m^2}.
\end{equation}
$H^{(1)}_A$ is the so-called {\sl Weyl} pseudo-differential operator defined with 
``mid-point prescription" treated in Ichinose--Tamura [ITa-86], 
Ichinose [I2-88, 3-89, 6-95]. 
$H^{(2)}_A$ is a modification of $H^{(1)}_A$ given 
by Iftimie--M\u{a}ntoiu--Purice [IfMp1-07, 2-08,3-10] with their other papers.
However, $H^{(3)}_A$ does not seem to be defined as
a pseudo-differential operator corresponding to a certain tractable symbol.
Indeed, so long as it is defined through Fourier and inverse-Fourier tansforms,
the candiadte of its symbol will not be $\sqrt{(\xi-A(x))^2+m^2}$ of (1.1).
The last $H^{(3)}$ is used, for instance, to study ``stability of matter" 
in relativistic quantum mechanics in Lieb--Seiringer [LSe-10].
Needles to say, we can show that these three relativistic Schr\"odinger 
operators $H^{(1)}$, $H^{(2)}$ and $H^{(3)}$ define 
{\it selfadjoint} operators  in $L^2({\bf R}^d)$. 

Then, letting $H$ be one of the magnetic relativistic Schr\"odinger operators  
$H^{(1)}, \, H^{(2)}$, $H^{(3)}$ with 
$H_A^{(1)}, \, H_A^{(2)}$, $H_A^{(3)}$ in (1.2), (1.3), (1.4), consider the following
{\sl imaginary-time relativistic Schr\"odinger equation}, 
i.e. maybe called {\sl heat equation} for $H-m$:
\begin{equation}
 \displaystyle{\frac{\partial}{\partial t}} u(x,t) = -[H-m]u(x,t), 
                                 \quad t>0, \quad x \in {\bf R}^d.
\end{equation}
The solution of the Cauchy problem with initial data $u(x,0) = g(x)$ 
is given by the semigroup $u(x,t) = (e^{-t[H-m]}g)(x)$. 
We want to deal with 
 path integral representation for each $e^{-t[H^{(j)}-m]}g\, (j=1,2,3)$.
The path integral concerned is connected with the 
{\it L\'evy process} (e.g. [IkW2-81/89], [Sa2-99], [Ap-04/09]) on the space 
$D_x := D_x([0,\infty)\rightarrow {\bf R}^d)$ 
dependent on each $x \in {\bf R}^d$, of the {\it ``c\`adlag} paths", 
i.e. right-continuous paths 
 $X: \![0,\infty)\! \ni s \mapsto X(s) \in \! {\bf R}^d$ 
having left-hand limits and satisfying $X(0)\!=\!x$. 
As our probability space $(\Omega, P)$ which is a pair of space $\Omega$ and 
probability $P$, though here and below not mentioning a $\sigma$-algebra 
on $\Omega$, we take a pair $(D_x, \lambda_x)$ of the path space $D_x$ 
and the associated path space measure $\lambda_x$ on $D_x$,
a probability measure whose characteristic function is given by 
\begin{equation}%
e^{-t[\sqrt{\xi^2+m^2}-m]} 
 = \int_{D_x([0,\infty)\rightarrow {\bf R}^d)} 
  e^{i (X(t)-x)\cdot\xi}d\lambda_x(X), \quad t\geq 0, 
\quad \xi \in {\bf R}^d.
\end{equation}
We will suppress use of the word ``random variable" $\omega \in \Omega$.

The aim of this article is to make review, mainly from our results, first on
some properties of these three magnetic Schr\"odinger operators as 
selfadjoint operators, for instance, 
that they are different in general from one another but to coincide 
when the vector potential $A(x)$ is linear in $x$, in particular, 
in the case of constant magnetic fields,
and bounded from below by the same greatest lower bound, with study
of whether they are gauge-covariant, 
mainly based on [I2-88, 3-89, 4-92, 8-12], [ITs1-92], [IIw-95], 
and next on Feyman--Kac--It\^o--type path integral representations 
for their respective {\sl imaginary-time} unitary groups, 
i.e. {\sl real-time} semigroups mainly based on [ITa-86], [I7-95], 
[HILo1-12, 2-12]. 
It will be of some interest to collect them 
in one place to observe how they look like and different, though 
all the three are bascally connected with the L\'evy process.

In Section 2 we give precise definition of the three magnetic 
relativistic Schr\"odinger operators and in Section 3 more 
general definition, studying their properties. 
In Section 4 path integral representaions for the semigroups for 
these three magnetic relativistic Schr\"odinger operators are given
accompanied with arguments about heuristic derivation. 
At the end a summary is given so as to be able 
to compare the three path integral formulas obtained.

The content of this article is an expanded version of the lecture with 
almost the same title given by the author at the International Conference 
on ``Partial Differential Equations and Spectral Theory" organized by
M. Demuth, B.-W. Schulze and I. Witt, in Goslar, Germany, 
August 31--September 6, 2008. A brief note with condensed content on the subject 
with sketch of proofs also is written in [I9-12] and 
a rather informal introductory paper of expositary character in [I10-12]. 
For one of the recent references on the related subjects we refer to [LoHB-11].
I hope the present work will give a little more 
extensive survey to give reviews.

\section{Three magnetic relativistic Schr\"odinger operators}

In Section 1, we introduced, though rather roughly, 
the three magnetic relativistic Schr\"odinger operators  
$H^{(1)} = H^{(1)}_A + V$, 
$H^{(2)} = H^{(2)}_A + V$, $H^{(3)}= H^{(3)}_A + V$ corresponding to 
the classical relativistic Hamiltonian symbol (1.1). 
In this Section we are going to give more 
unambiguous defintions of them and study their properties.
The difference lies in how to define their first terms on the right,  
kinetic energy operators $H^{(1)}_A$, $H^{(2)}_A $, $H^{(3)}_A$,  
corresponding to the part $\sqrt{(\xi-A(x))^2 +m^2}$ of the symbol (1.1). 

\subsection{Their definition and difference}

For simplicity, it is assumed here as in Section 1 that the 
vector potential $A: {\bf R}^d \rightarrow {\bf R}^d$ is a $C^{\infty}$
function and the scalar potential $V: {\bf R}^d \rightarrow {\bf R}$ is a 
function {\it bounded below}. The space of the $C^{\infty}$ functions 
with compact support and the space of rapidly decreasing $C^{\infty}$ 
functions in ${\bf R}^d$ are denoted respectively by $C_0^{\infty}({\bf R}^d)$ 
and ${\cal S}({\bf R}^d)$.

The definition of  $H^{(1)}_A$, $H^{(2)}_A$ as pseudo-differential operators 
in (1.2), (1.3) needs the concept of {\it oscillatory integrals}. 
If the symbol $a(\eta,y)$ satisfies 
for some $m_0 \in {\bf Z}$ and $\tau_0 \geq 0$ that 
for any multi-indices $\alpha :=(\alpha_1,\dots,\alpha_d)$,   
$\beta :=(\beta_1,\dots,\beta_d)$ of nonnegative integers
there exist constants $C_{\alpha \beta}$ such that  
$$|\partial_{\eta}^{\alpha}\partial_y^{\beta} a(\eta,y)| 
\leq C_{\alpha \beta}(1+|\eta|^2)^{m_0/2}(1+|y|^2)^{\tau_0/2}, 
$$
then the oscillatory integral (e.g. [Ku-74, Theorem 6.4, p.47])
\begin{equation}
\hbox{\rm Os--}\!\int\int_{{\bf R}^d \times {\bf R}^d}
e^{-iy\cdot \eta}a(\eta,y) dyd\eta
 := \lim_{\varepsilon \rightarrow 0+} \int\int_{{\bf R}^d \times {\bf R}^d} 
e^{-iy\cdot \eta} \chi(\varepsilon\eta,\varepsilon y)a(\eta,y) dyd\eta
\end{equation}
exists, where $\chi(\eta,y)$ is any cutoff function in  
${\cal S}({\bf R}^d \times {\bf R}^d)$  such that $\chi(0,0)=1$. 
The existence of the limit on the right-hand side of (2.1) is 
independent of the choice of cutoff functions $\chi$, and 
shown by integration by parts as follows. First note that
$$(-i\partial_y)^{\beta}e^{-iy\cdot\eta}= (-\eta)^{\beta}e^{-iy\cdot\eta},\quad
(-i\partial_{\eta})^{\alpha}e^{-iy\cdot\eta}= (-y)^{\alpha}e^{-iy\cdot\eta},
$$
so that
$$\langle\eta\rangle^{-2l} \langle -i\partial_y\rangle^{2l}
e^{-iy\cdot\eta}= e^{-iy\cdot\eta},\quad 
\langle y\rangle^{-2l'} \langle -i\partial_{\eta}\rangle^{2l'}
e^{-iy\cdot\eta}= e^{-iy\cdot\eta}, 
$$
where
$$
 \langle\eta\rangle = (1+|\eta|^2)^{1/2},\,\, 
 \langle y \rangle =(1+|y|^2)^{1/2},\,\,
 \langle -i\partial_y\rangle^2 = (1-\Delta_y), \,\,
 \langle -i\partial_{\eta}\rangle^2 = (1-\Delta_{\eta}).
$$
Then the above note with integration by parts shows the integral 
before the limit $\varepsilon \rightarrow 0+$ taken is equal to 
\begin{eqnarray*}
&& \int\int e^{-iy\cdot \eta} 
 \langle \eta\rangle^{-2l}\langle-i\partial_y\rangle^{2l}
 \big(\chi(\varepsilon\eta,\varepsilon y)a(\eta,y)\big)dyd\eta\\
&=& \int\int e^{-iy\cdot \eta} 
 \langle y\rangle^{-2l'}\langle-i\partial_{\eta}\rangle^{2l'}
 \big[\langle \eta\rangle^{-2l}\langle-i\partial_y\rangle^{2l}
 \big(\chi(\varepsilon\eta,\varepsilon y)a(\eta,y)\big)\big]dyd\eta\,.
\end{eqnarray*}
In the integral on the right above, if the positive integers $l$ and $l'$ 
are so taken that $-2l + m_0 <-d, \,\,-2l' +\tau_0 < -d$, then 
$
 \langle y\rangle^{-2l'}\langle-i\partial_{\eta}\rangle^{2l'}
 \big(\langle \eta\rangle^{-2l}\langle-i\partial_y\rangle^{2l} a(\eta,y)\big)
$
becomes integrable on ${\bf R}^d \times {\bf R}^d$. Therefore
taking the limit $\varepsilon \rightarrow 0+$, we see
by the Lebesgue dominated convergence theorem this integral 
converges to 
$$\int\int
  \langle y\rangle^{-2l'}\langle-i\partial_{\eta}\rangle^{2l'}
 \big(\langle \eta\rangle^{-2l}\langle-i\partial_y\rangle^{2l} a(\eta,y)\big) 
dyd\eta,
$$
which implies  existence of the integral 
$\hbox{\rm Os-}\!\int\int e^{-iy\cdot \eta}a(\eta,y) dyd\eta$, 
showing existence of the oscillatory integral (2.1).

\medskip
For $H^{(1)}_A$ in (1.2) we have the following proposition. 
\begin{prp} 
Let $m \geq 0$. If $A(x)$ is in $C^{\infty}({\bf R}^d; {\bf R}^d)$ and 
satisfies for some $\tau \geq 0$ that
\begin{equation}  
|\partial_x^{\beta}A(x)| \leq C_{\beta} \langle x \rangle^{\tau}
\end{equation}
for any multi-indices $\beta$ with constants $C_{\beta}$, 
then for $f$ in $C_0^{\infty}({\bf R}^d)$
the Weyl pseudo-differential operator $H^{(1)}_A$ in (1.2) exists 
as an oscillatory integral and further is equal to a second expression;
namely, one has
\begin{eqnarray}
(H^{(1)}_A f)(x) 
\!\!&\!\!=\!\!&\!\! \frac1{(2\pi)^{d}} \hbox{\rm Os--}\! 
  \displaystyle{\int\int}_{{\bf R}^d\times {\bf R}^d} 
   e^{i(x-y)\cdot\xi}\sqrt{\Big(\xi-A\big(\frac{x+y}{2}\big)\Big)^2 +m^2}
   \,f(y) dyd\xi,\qquad \\
&\!\!=\!\!&\!\! \frac1{(2\pi)^{d}} \hbox{\rm Os--}\! 
  \displaystyle{\int\int}_{{\bf R}^d\times {\bf R}^d}
   e^{i(x-y)\cdot\big(\xi+A(\frac{x+y}{2})\big)}
      \sqrt{\xi^2 +m^2} \,f(y) dyd\xi.
\end{eqnarray}
\end{prp}

\bigskip
{\it Proof}. 
We give only a sketch of the proof, dividing into the two cases $m>0$ and $m=0$,
where note that in the former case the symbol 
$\big((\xi-A(x))^2 +m^2\big)^{1/2}$ has no singularity, but in the latter case 
singularity on the set 
$\{(\xi,x) \in {\bf R}^d\times {\bf R}^d \,;\,|\xi-A(x)|=0\}$.

(a) {\sl The case $m>0$}. First we treat the oscillatory integral (2.3). 
Note that
$|\partial_{\xi}^{\alpha}\partial_x^{\beta}\big((\xi-A(x))^2 +m^2\big)^{1/2}| 
\leq C_{\alpha \beta}
\langle \xi \rangle\langle x \rangle^{\tau}$, and hence
$|\partial_{\xi}^{\alpha}\partial_x^{\beta}
\big[\big((\xi-A(x))^2 +m^2\big)^{1/2}f(x)\big]|\leq C_{\alpha \beta}
\langle \xi \rangle\langle x \rangle^{\tau}$.

Since $f$ is taken from $C_0^{\infty}({\bf R}^d)$, we may take 
a cutoff function which is only dependent on the variable $\xi$ 
but not $y$, i.e. $\chi \in {\cal S}({\bf R}^d)$ with $\chi(0)=1$. 
Then we have by integration by parts as seen before this proposition,
\begin{eqnarray}
(H_A^{(1)}f)(x)
\!\!&\!=\!&\!\! \lim_{\varepsilon \rightarrow 0+}
\frac1{(2\pi)^{d}}\int\int_{{\bf R}^d\times {\bf R}^d} 
 e^{i(x-y)\cdot\xi}
\chi(\varepsilon \xi) 
\sqrt{\Big(\xi-A\big(\frac{x+y}{2}\big)\Big)^2 +m^2}\,f(y) dyd\xi 
                                          \nonumber\\
\!\!&\!=\!&\!\! \lim_{\varepsilon \rightarrow 0+}
\frac1{(2\pi)^{d}}\int\int_{{\bf R}^d\times {\bf R}^d} 
 e^{i(x-y)\cdot\xi} \nonumber\\
&&\qquad \times
\langle\xi\rangle^{-2l} \langle -i\partial_y\rangle^{2l}\Big(
\chi(\varepsilon \xi) 
\sqrt{\Big(\xi-A\big(\frac{x+y}{2}\big)\Big)^2 +m^2}\,f(y)\Big) dyd\xi\,.
\end{eqnarray}
If we take $l$ sufficiently large, 
$
 \langle\xi\rangle^{-2l} \langle -i\partial_y\rangle^{2l}\big(
\sqrt{\big(\xi-A(\frac{x+y}{2})\big)^2 +m^2}\,f(y)\big)
$
becomes integrable on ${\bf R}^d\times {\bf R}^d$ for fixed $x$, so that 
as $\varepsilon \rightarrow 0+$, 
by the Lebesgue dominated convergence theorem  we have
\begin{eqnarray}
(H_A^{(1)}f)(x)
\!\!&\!=\!&\!\! \frac1{(2\pi)^{d}}\int\int_{{\bf R}^d\times {\bf R}^d} 
 e^{i(x-y)\cdot\xi} \nonumber\\
&&\quad \times
\langle\xi\rangle^{-2l} \langle -i\partial_y\rangle^{2l}\Big(
\sqrt{\Big(\xi-A\big(\frac{x+y}{2}\big)\Big)^2 +m^2}\,f(y)\Big) dyd\xi\,.
\end{eqnarray}
This proves existence of the second oscillatory integral (2.3) for $H_A^{(1)}$
when $m>0$. Similarly, we can show existence of the second oscillatory 
integral (2.4). This shows the first part of the proposition. 

To show the second part, i.e. coincidence of the two expressions (2.3)
and (2.4), first suppose that $A(x)$ is $C_0^{\infty}$.
In the integral of the second member of (2.5), we make the change 
of variables: 
$\xi =\xi'+A\big(\frac{x+y'}{2}\big),\, y=y'$, where we note the Jacobian
$\Big|\frac{\partial(\xi,y)}{\partial(\xi',y')}\Big| = 1$.
Then it (= the right-hand side of (2.5)) is equal, 
with $\xi',\,\, y'$ rewritten as $\xi,\,\, y$ again, to  
\begin{eqnarray*}
&&\lim_{\varepsilon \rightarrow 0+}\int\int_{{\bf R}^d\times {\bf R}^d} 
 e^{i(x-y)\cdot\big(\xi+A(\frac{x+y}{2})\big)}
\chi\Big(\varepsilon \big(\xi+A(\frac{x+y}{2})\big)\Big)\sqrt{\xi^2+m^2}
  \,f(y) dyd\xi \qquad \qquad \\
\!\!&\!=\!&\!\! \lim_{\varepsilon \rightarrow 0+}\int\int
 e^{i(x-y)\cdot\xi}
\langle \xi\rangle^{-2l}\langle -i\partial_y\rangle^{2l}\Big[
\chi\Big(\varepsilon \big(\xi+A(\frac{x+y}{2})\big)\Big)\\
 && \qquad \qquad\qquad \qquad\qquad \qquad\qquad \qquad 
 \times 
  \sqrt{\xi^2+m^2}\,e^{i(x-y)\cdot A(\frac{x+y}{2})}f(y) \Big]
   dyd\xi\,,
\end{eqnarray*}
where we have integrated by parts when passing from 
the left-hand side to the right.
Note that, since the factor 
$\langle \xi\rangle^{-2l}\langle -i\partial_y\rangle^{2l}\Big(
   \sqrt{\xi^2+m^2}\,e^{i(x-y)\cdot A(\frac{x+y}{2})}f(y) \Big) 
$ 
in the integrand on the right-hand side  
is integrable on ${\bf R}^d\times {\bf R}^d$  for $l$ sufficiently large
with $x$ fixed, 
we can take the limit $\varepsilon \rightarrow 0+$. So the right-hand side
turns out to be equal to 
\begin{eqnarray}
&&\int\int  e^{i(x-y)\cdot\xi}
\langle \xi\rangle^{-2l}\langle -i\partial_y\rangle^{2l}\Big(
  \sqrt{\xi^2+m^2}\,e^{i(x-y)\cdot A(\frac{x+y}{2})}f(y) \Big)
   dyd\xi \nonumber\\
\!\!&\!=\,\!&\!\! \lim_{\varepsilon \rightarrow 0+}\int\int  e^{i(x-y)\cdot\xi}
 \langle \xi\rangle^{-2l}\langle -i\partial_y\rangle^{2l}\Big(
\chi(\varepsilon \xi)  
\sqrt{\xi^2+m^2}\,e^{i(x-y)\cdot A(\frac{x+y}{2})}f(y) \Big)dyd\xi
                                              \nonumber\\
\!\!&\!=\,\!&\!\! \lim_{\varepsilon \rightarrow 0+}
   \int\int  e^{i(x-y)\cdot\xi}
 \chi(\varepsilon \xi) \sqrt{\xi^2+m^2}\,e^{i(x-y)\cdot A(\frac{x+y}{2})}
   f(y)dyd\xi \nonumber\\
\!\!&\!=:\!&\! \hbox{\rm Os--}\! 
  \int\int_{{\bf R}^d\times {\bf R}^d}
   e^{i(x-y)\cdot\big(\xi+A(\frac{x+y}{2})\big)}
      \sqrt{\xi^2 +m^2} \,f(y) dyd\xi.
\end{eqnarray}
Here the second equality is due to integration by parts.
This shows coincidence of (2.3) and (2.4) for 
$A \in C_0^{\infty}({\bf R}^d; {\bf R}^d)$.

Next, we come to the general case where 
$A \in C^{\infty}({\bf R}^d;\,{\bf R}^d)$
satisfies (2.2). For this $A(x)$ there exists a sequence 
$\{A_n(x)\}_{n=1}^{\infty} \subset C_0^{\infty}({\bf R}^d; {\bf R}^d)$ 
which converges to $A(x)$ in the topology of $C^{\infty}({\bf R}^d;\,{\bf R}^d)$, 
i.e. the $A_n(x)$, together with all their derivatives, converge to $A(x)$ 
as $n\rightarrow \infty$ uniformly on every compact subset of ${\bf R}^d$.
Then we have seen above the coincidence of the two expressions (2.3) and (2.4) 
for the Weyl pseudo-differential operatoret $H_{A_n}^{(1)}$ 
corresponding to the symbol $\big((\xi-A_n(y))^2 +m^2\big)^{1/2}$.
Therefore, observing (2.6) and (2.7) with $A_n$  in place of $A$, 
we obtain 
\begin{eqnarray*}
&&\int\int 
 e^{i(x-y)\cdot\xi} 
\langle\xi\rangle^{-2l} \langle -i\partial_y\rangle^{2l}\Big(
\sqrt{\Big(\xi-A_n\big(\frac{x+y}{2}\big)\Big)^2 +m^2}\,f(y)\Big) dyd\xi\\
&=& (2\pi)^{d}(H_{A_n}^{(1)}f)(x)\\
&=&\int\int  e^{i(x-y)\cdot\xi}
\langle \xi\rangle^{-2l}\langle -i\partial_y\rangle^{2l}\Big(
  \sqrt{\xi^2+m^2}\,e^{i(x-y)\cdot A_n(\frac{x+y}{2})}f(y) \Big)
   dyd\xi\,.
\end{eqnarray*}
Then we see with the Lebesgue convergence theorem that
as $n \rightarrow \infty$,
the first member and the third converge to (2.3) and (2.4), respectively,
showing the coincidence of (2.3) and (2.4) in the general case.
 
\medskip
(b) {\sl The case $m=0$}. For our cutoff function $\chi(\xi)$, take it from   
$C_0^{\infty}({\bf R}^d)$ and require further rotational symmetricity such that
$0\leq \chi(\xi) \leq 1$ for all $\xi \in {\bf R}^d$ 
and $\chi(\xi) =1$ for $|\xi|\leq \frac12;\,\, =0$ for $|\xi|\geq 1$. 
Put $\chi_n(\xi) = \chi(\xi/n)$ for positive integer $n$.
Then split the symbol $|\xi-A(x)|$ into a sum of two terms: 
$|\xi-A(x)| = h_1(\xi,x)  + h_2(\xi,x) $, 
$$
  h_1(\xi,x) = \chi_n(\xi-A(x))|\xi-A(x)|, 
\qquad h_2(\xi,x) = [1-\chi_n(\xi-A(x))]|\xi-A(x)|.
$$
Although then the symbol $h_1(\xi,x)$ has singularity, the corresponding 
Weyl pseudo-differential operator can define a bounded operator on 
$L^2({\bf R}^d)$ well, and so there is no problem. The one corresponding to 
the symbol $h_2(\xi,x)$, which has no more singularity, 
is a pseudo-differential operator defined by oscillatory integral, to which
the method in the case (a) above will apply.
This ends the proof of Proposition 2.1. \qed 

\medskip
For $H^{(2)}_A$ in (1.3), we can show 
the following proposition in the same way as Proposition 2.1.
\begin{prp} 
Under the same hypothesis for $A(x)$ as in Proposition 2.1, 
for $f$ in $C_0^{\infty}({\bf R}^d)$ 
the pseudo-differential operator  
$H^{(2)}_A$ in (1.3) exists as an oscillatory integral
and further is equal to a second expression; namely, one has
\begin{eqnarray}
(H^{(2)}_A f)(x)
\!\!&\!=\!&\!\! \frac1{(2\pi)^{d}} \hbox{\rm Os--}\! 
  \displaystyle{\int\int}_{{\bf R}^d\times {\bf R}^d} 
   e^{i(x-y)\cdot\xi} \nonumber \\
&&\qquad\quad \times \sqrt{\Big(
\xi-\int_0^1 A((1-\theta)x+\theta y)d\theta\Big)^2 +m^2}
   \,f(y) dyd\xi,\qquad  \\
\!\!&\!=\!&\!\! \frac1{(2\pi)^{d}} \hbox{\rm Os--}\! 
  \displaystyle{\int\int}_{{\bf R}^d\times {\bf R}^d}
   e^{i(x-y)\cdot\big(\xi+\int_0^1 A((1-\theta)x+\theta y)d\theta\big)}
                      \nonumber \\
&&\qquad\qquad\qquad\qquad\qquad\qquad\qquad\quad
 \times
   \sqrt{\xi^2 +m^2} \,f(y) dyd\xi.
\end{eqnarray}
\end{prp}

\medskip
In the following, let us gather here, for our three magnetic relativistic 
Schr\"odinger operators $H^{(1)}_A$,  $H^{(2)}_A$,  $H^{(3)}_A$,
all their defintion for up to the present. The most  general definition 
will be given in Section 3.

\begin{dfn} 
{\rm  For $A \in C^{\infty}({\bf R}^d; {\bf R}^d)$ satisfying condition 
(2.2), $H^{(1)}_A$ is defined as the pseudo-differential operators 
(2.3) and/or (2.4).
}
\end{dfn}

\begin{dfn} 
{\rm  For $A \in C^{\infty}({\bf R}^d; {\bf R}^d))$ satisfying  condition 
(2.2), $H^{(2)}_A$ is defined as the pseudo-differential operators 
(2.8) and/or (2.9).
}
\end{dfn}

The definition of $H^{(3)}_A$ encounters a situation totally different 
from the previous $H^{(1)}_A$ and $H^{(2)}_A$ case. We 
need nonnegative selfadjointness of the operator
$(-\nabla-A(x))^2$ in $L^2({\bf R}^d)$, which can be considered as 
nothing but the nonrelativistic magnetic Schr\"odinger operator 
for a particle with mass $m=\frac12$. Of course, if 
$A \in C^{\infty}({\bf R}^d; {\bf R}^d)$,  $(-\nabla-A(x))^2$ becomes
a nonnegative, selfadjoint operator. 
However, more generally, it is shown by Kato and Simon 
(see [CFGKS, pp.8--10]) that 
if $A \in L^2_{{\rm loc}}({\bf R}^d; {\bf R}^d)$,
 $C_0^{\infty}({\bf R}^d)$ is a form core for the quadratic form 
for $(-\nabla-A(x))^2$ in $L^2({\bf R}^d)$, so that by the well 
known argument (e..g. [Kat-76, VI, \S 2, Theorems 2.1, 2.6, pp.322--323]) 
there exists a unique nonnnegative, selfadjoint operator 
in $L^2({\bf R}^d)$ associated with this quadratic form  with form domain 
$\{u \in L^2({\bf R}^d)\,;\, (-\nabla-A(x)) u \in L^2({\bf R}^d\}$. 
One may take it as $(-\nabla-A(x))^2$.  
Then its square root $\sqrt{(-i\nabla-A(x))^2+m^2}$ exists as a 
nonnegative, selfadjoint operator in $L^2({\bf R}^d)$. 
This give the following definition for $H^{(3)}_A$.

\begin{dfn} 
{\rm If For $A \in L^2_{{\rm loc}}({\bf R}^d; {\bf R}^d)$, 
$H^{(3)}_A$ is defined as the square root :
\begin{equation}%
H^{(3)}_A := \sqrt{(-i\nabla-A(x))^2+m^2}
\end{equation}
of the nonnegative selfadjoint operator $(-i\nabla-A(x))^2+m^2$ 
in $L^2({\bf R}^d)$.
}
\end{dfn}
We note that this $H^{(3)}_A$ does not seem to be defined as
a pseudo-differential operator corresponding to a certain tractable symbol.
So long as pseudo-differential operators are defined through 
Fourier and inverse-Fourier tansforms,
the candidate of its symbol will not be $\sqrt{(\xi-A(x))^2+m^2}$.
The $H_A^{(3)}$ is used, for instance, to study ``stability of matter" 
in relativistic quantum mechanics in Lieb--Seiringer [LSei-10].
An kinetic energy inequality in the presence of the vector potential
for the relativistic Schr\"odinger operators
$H_A^{(1)}$ and $H_A^{(3)}$ as well as the nonrelativistic 
Schr\"odinger operator $(2m)^{-1}(-i\nabla -A(x))^2$ was given in [I6-93].

\medskip
Needles to say, we can show that not only $H_A^{(3)}$ but also $H_A^{(1)}$ and 
$H_A^{(2)}$ define selfadjoint operators 
 in $L^2({\bf R}^d)$. They are in general different from one another 
but coincide with one another if $A(x)$ is linear in $x$. 
We observe these facts in the following.

\begin{prp} 
$H_A^{(1)}$, $H_A^{(2)}$ and $H_A^{(3)}$ are in general different.
\end{prp}

{\it Proof}. 
First, one has $H_A^{(1)} \not= H_A^{(2)}$ for general vector potentials $A$, 
because we have
$$
 A\big(\frac{x+y}{2}\big) \not= \int_0^1 A(x+\theta(y-x)) d\theta.
$$
Indeed, for instance, for $d=3$, taking 
$A(x) \equiv (A_1(x),A_2(x),A_3(x))= (0,0,x_3^2)$, we have
\begin{eqnarray*}
\int_0^1 A_3(x+\theta(y-x)) d\theta 
&=& \int_0^1 (x_3+\theta(y_3-x_3))^2 d\theta 
= \frac{x_3^2+x_3y_3+y_3^2}{3} \qquad \\
&\not=& (\frac{x_3+y_3}{2})^2 = A_3\big(\frac{x+y}{2}\big).
\end{eqnarray*}

Next, to see that $H_A^{(1)} \not= H_A^{(3)}$ and $H_A^{(2)} \not= H_A^{(3)}$, 
one needs to show (e.g. [Ho-85, Section 18.5, p.150--152]),  
for some $g \in C_0^{\infty}({\bf R}^3)$, respectively, that 
\begin{eqnarray*}
&\!\!&\!\!\frac1{(\!2\pi\!)^{6}}\!\int\!\int\int\!\int\! 
e^{i(x-z)\cdot\zeta+i(z-y)\cdot\eta}
\Big[\Big(\zeta-A\big(\frac{x+z}2\big)\Big)^2+m^2\Big]^{1/2}\\
&&\times
  \Big[\Big(\eta-A\big(\frac{z+y}2\big)\Big)^2+m^2\Big]^{1/2}g(y)
 dz d\zeta dy d\eta
\not= [(-\nabla-A(x))^2+m^2]g(x),
\end{eqnarray*}
and that 
\begin{eqnarray*}
&\!\!&\!\!\frac1{(\!2\pi\!)^{6}}\!\int\!\int\int\!\int\! 
e^{i(x-z)\cdot\zeta+i(z-y)\cdot\eta}
\Big[\Big(\zeta-\!\int_0^1 A\big(x+\theta(z-x)\big)d\theta
    \Big)^2+m^2\Big]^{1/2}\\
&&\times
\Big[\Big(\eta-\!\int_0^1 A\big(z+\theta(y-z)\big)\Big)^2+m^2\Big]^{1/2}g(y)
 dz d\zeta dy d\eta
\not= [(-\nabla-A(x))^2+m^2]g(x).
\end{eqnarray*}
Here the integrals with respect to the space variables above and below 
are oscillatory integrals.

The former for $H_A^{(1)}$ was shown by Umeda--Nagase 
[UNa-93, Section 7, p.851]. Indeed,
putting $p(x,\xi) := \sqrt{(\xi-A(x))^2+m^2}$, they verified the (Weyl) symbol
$(p\circ p)(x,\xi)$ of $(H_A^{(1)})^2$ satisfy (see 
[UNa-93, Lemma 6.3, p.846]; [Ho-pp.151--152]) that
\begin{eqnarray*}
(p\circ p)(x,\xi)
\!&\!=\!& \!\frac1{(\!2\pi\!)^{6}}\!\int\!\int\int\!\int\! 
  e^{-i(y\cdot\eta+z\cdot\zeta)}
  p(x+\frac{z}2,\xi-\eta) p(x-\frac{y}2,\xi-\eta)dy d\eta dz d\zeta \\
\!&\!=\!& \!\frac1{(\!2\pi\!)^{6}}\!\int\!\int\int\!\int\! 
e^{i(x-z+y/2)\cdot(\zeta-\xi)+ i(z-x+y/2)\cdot(\eta-\xi)}\\
&&\qquad\qquad\qquad \times
p\big(\frac{x+z+y/2}2, \zeta\big) p\big(\frac{x+z-y/2}2, \eta\big)
 dz d\zeta dy d\eta \\
\!&\!\not=\!&\! (\xi-A(x))^2 +m^2.
\end{eqnarray*}

The latter for $H_A^{(2)}$ will be shown in a similar way.
\qed

\bigskip\noindent
\begin{thm} 
If $A(x)$ is linear in $x$, i.e. if 
$A(x) = \dot{A}\cdot x\,$ with $\dot{A}$ being any $d\times d$ 
real symmetric constant matrix, then
$H_A^{(1)},\,H_A^{(2)}$ and $H_A^{(3)}$ coincide. In particular, 
this holds for constant magnetic fields with $d=3$,
i.e.  when $\nabla \times A(x)$ is constant.
\end{thm}

{\it Proof}.  Suppose $A(x) = \dot{A}\cdot x$. 
First, we see that $H_A^{(1)} = H_A^{(2)}$ because we have
\begin{eqnarray*}
\int_0^1 {A}(\theta x +(1-\theta)y) d\theta
&=&\int_0^1 \dot{A}\cdot(\theta x +(1-\theta)y) d\theta
= \int_0^1  \dot{A}\cdot\big(y+\theta (x-y)\big)d\theta\nonumber\\ 
&=& \dot{A}\cdot \frac{x+y}{2}
= A\big(\frac{x+y}{2}\big),
\end{eqnarray*}
which turns out to be ``midpoint prescription"
to yield the Weyl quantization.

To see that they also concide with $H_A^{(3)}$, we need to show that 
$(H_A^{(1)})^2 = (-i\nabla -A(x))^2+m^2$.
To do so, let
$f$ $\in C_0^{\infty}({\bf R}^d)$ and 
note $(\dot{A})^T= \dot{A}$, then we have, 
with integrals as oscillatory integrals,
\begin{eqnarray*}
&&\! \big((H_A^{(1)})^2 f\big)(x)\\
&\!=\!&\! \frac1{(\!2\pi\!)^{2d}}
\!\int\!\int\int\!\int\!
e^{i(x-z)\cdot(\eta+\dot{A} \frac{x+z}{2})
          +i(z-y)\cdot(\xi+\dot{A} \frac{z+y}{2})}
\!\sqrt{\eta^2+m^2}\sqrt{\xi^2+m^2}\!f(y)dzd\eta dyd\xi\\
&\!=\!&\!\frac1{(\!2\pi\!)^{2d}}
\!\int\!\int\!\int\!\int\!
   e^{iz\cdot(-\eta+\xi)}
    e^{i[x\cdot(\eta+\dot{A}\frac{x}{2})-y\cdot(\xi+\dot{A}\frac{y}{2})]}
\!\sqrt{\eta^2+m^2}\sqrt{\xi^2+m^2}\!f(y)dzd\eta dy\xi\\
&\!=\!&\frac1{(2\pi)^{d}}\int\!\int\!\int\! \delta(-\eta+\xi)
    e^{i[x\cdot(\eta+\dot{A}\frac{x}{2})-y\cdot(\xi+\dot{A}\frac{y}{2})]}
\sqrt{\eta^2+m^2}\sqrt{\xi^2+m^2}f(y)\,d\eta dyd\xi\,. 
\end{eqnarray*}
Hence
\begin{eqnarray*}
\big((H_A^{(1)})^2 f\big)(x)   
&\!=\!&\frac1{(2\pi)^{d}}\int\!\int
    e^{i(x-y)\cdot\xi}e^{i\frac12(x\cdot\dot{A}x -y\cdot\dot{A}y)}
    (\xi^2+m^2)f(y)\, dyd\xi\\ 
&\!=\!& \frac1{(2\pi)^{d}}
  \int\!\int e^{i(x-y)\cdot(\xi+\dot{A}\frac{x+y}{2})}
  (\xi^2+m^2)f(y)\,dyd\xi\\
&\!=\!& \frac1{(2\pi)^{d}}\int\!\!\int 
 e^{i(x-y)\cdot(\xi+A(\frac{x+y}{2}))}
  (\xi^2+m^2)f(y)\,dyd\xi\\ 
&\!=\!& \frac1{(2\pi)^{d}}\int\!\int e^{i(x-y)\cdot\xi}
  \big[\big(\xi-A\big(\frac{x+y}{2}\big)\big)^2+m^2\big]f(y)\,dyd\xi\,. 
\end{eqnarray*}
The last equality is due to the fact that 
symbol $(\xi-A(x))^2+m^2$ is polynomial of $\xi$, so that
the corresponding Weyl pseudo-differential operator is equal to
$(-i\nabla -A(x))^2+m^2$. 
\qed

\subsection{Gauge-covariant or not} 

Among these three magnetic relativistic Schr\"odinger operators 
$H_A^{(1)}$, $H_A^{(2)}$ and $H_A^{(3)}$, 
the Weyl quantized one like $H_A^{(1)}$ 
(in general, the Weyl pseudo-differential operator) is 
{\sl compatible well with path integral} (e.g. Mizrahi [M-78]). 
But the pity is that, for general vector potential $A(x)$, 
$H_A^{(1)}$ (and so $H^{(1)}$) is not generally covariant 
under gauge transformation, namely, 
there exists a real-valued function
$\varphi(x)$ for which it fails to hold that
$H^{(1)}_{A+\nabla \varphi} = e^{i\varphi}H^{(1)}_{A}e^{-i\varphi}$.

However, $H_A^{(2)}$ (and so $H^{(2)}$) and $H_A^{(3)}$ (and so $H^{(3)}$)
are gauge-covariant, though these three are not in general equal
as seen in Proposition 2.6.
The gauge-covariance of the modified $H^{(2)}_A$ in contrast to $H^{(1)}_A$
in Ichinose--Tamura [ITa-86] was emphasized in Iftimie--M\u{a}ntoiu--Purice 
[IfMP1-07, 2-08, 3-10]. There, in particular, in [IfMP1-07], they also compared 
our three magnetic Schr\"odinger operators to observe the following facts.

\begin{prp}
$H_A^{(2)}$ and $H_A^{(3)}$ are covariant under gauge transformation, 
i.e. it holds for $j=2,3$ that 
$\,\,H^{(j)}_{A+\nabla \varphi} = e^{i\varphi}H^{(j)}_{A}e^{-i\varphi}$ 
 for every $\varphi \in {\cal S}({\bf R}^d)$.
But $H_A^{(1)}$ is in general not covariant under gauge transformation.
\end{prp}

\medskip
{\it Poof}. First, to see the assertion for 
$H_A^{(3)}= \sqrt{(-i\nabla-A(x))^2+m^2}$, put 
$K_A = (-i\nabla-A(x))^2 +m^2$, so that $H_A^{(3)} = {K_A}^{1/2}$. As 
$(-i\nabla-A(x))^2$ is a nonrelativistic magnetic Schr\"odinger operator
with mass $\frac12$, being a nonnegative selfadjoint operator on $L^2({\bf R}^d)$,
and gauge-covariant,  so is $K_A$. Therefore it satisfies
$K_{A+\nabla \varphi} = e^{i\varphi}K_A e^{-i\varphi}$ for every $\varphi(x)$.
It follows that 
$({K_{A+\nabla \varphi}}^{1/2})^2 
= (e^{i\varphi}{K_A}^{1/2} e^{-i\varphi})(e^{i\varphi}{K_A}^{1/2}  e^{-i\varphi})$,
whence ${K_{A+\nabla \varphi}}^{1/2} = e^{i\varphi}{K_A}^{1/2} e^{-i\varphi}$,
because  $e^{i\varphi}{K_A}^{1/2} e^{-i\varphi}$ is also nonnegative 
selfadjoint.
This means that 
$H^{(3)}_{A+\nabla \varphi} = e^{i\varphi}H_A^{(3)}e^{-i\varphi}$,
i.e. $H^{(3)}_A$ is gauge-covariant.

Next, for $H_A^{(2)}$, by mean-value theorem 
\begin{eqnarray*}
\varphi(y) - \varphi(x) 
&=& \int_0^1 (y-x)\cdot(\nabla\varphi)(x+\theta (y-x))d\theta\\
&=& -\int_0^1 (x-y)\cdot(\nabla\varphi)((1-\theta)x+\theta y)d\theta.
\end{eqnarray*}
Hence
\begin{eqnarray*}
&&(H^{(2)}_A e^{-i\varphi}f)(x) \\
&\!=\!&\frac1{(2\pi)^{d}} \int\int_{{\bf R}^d\times {\bf R}^d}\, 
   e^{i(x-y)\cdot (\xi +\int_0^1 A((1-\theta)x+\theta y)d\theta)}\\
 &&\quad\qquad\qquad\times\, 
    \sqrt{\xi^2 +m^2}\,e^{-i\varphi(x)
  +i\int_0^1 (x-y)\cdot(\nabla\varphi)((1-\theta)x+\theta y)d\theta}
    f(y)\, dyd\xi \\
&\!=\!&\frac1{(2\pi)^{d}}
e^{-i\varphi(x)}\int\int_{{\bf R}^d\times {\bf R}^d}\, 
 e^{i(x-y)\cdot 
(\xi +\int_0^1 (A+\nabla \varphi)((1-\theta)x+\theta y)d\theta)}
   \sqrt{\xi^2 +m^2}\,f(y)\, dyd\xi \\
&\!=\!&
e^{-i\varphi(x)}(H^{(2)}_{A+\nabla\varphi}f)(x). 
\end{eqnarray*} 

Finally, we show non-gauge-invariance of $H_A^{(1)}$. To this end, 
we are going to use a second expression for $H^{(1)}_A$ as 
an integral operator to be given in the next section, (3.7) in Definition 3.7.
Then we show that it does not hold for all $\varphi$ that
$H^{(1)}_{A+\nabla\varphi} = e^{i\varphi}H_A^{(1)}e^{-i\varphi}$ or that,
 taking $A \equiv 0$. Indeed, suppose that 
\begin{eqnarray*}
&&\hbox{\rm p.v.}\!\int_{|y|>0} 
        [e^{-iy\cdot (\nabla\varphi)(x+\frac{y}{2})} f(x+y) -f(x)]\,n(dy)\\
&=& e^{i\varphi(x)}\hbox{\rm p.v.}\!\int_{|y|>0} 
        [(e^{-i\varphi}f)(x+y) -(e^{-i\varphi}f)(x)]\,n(dy) \\
&\equiv& \hbox{\rm p.v.}\!\int_{|y|>0} 
        [e^{-i(\varphi(x+y)-\varphi(x))}f(x+y) -f(x)]\,n(dy). 
\end{eqnarray*}
However, the second equality cannot hold, because it does not hold for all 
$\varphi$ that
 $\varphi(x+y) -\varphi(x) = y\cdot(\nabla\varphi)(x+\frac{y}{2})$.
\qed


\section{More general definition of magnetic relativistic Scr\"odinger 
operators and their selfadjointness}

In this section,
we want to give the most general definition of $H_A^{(1)}$, $H_A^{(2)}$ 
and $H_A^{(3)}$, which do {\sl not appeal to} the pseudo-differential operators.

\subsection{The most general definition of $H_A^{(1)}$, $H_A^{(2)}$ 
and $H_A^{(3)}$}  

First we concern $H_A^{(1)}$ and $H_A^{(2)}$. 
The starting point is the {\it L\'evy--Khinchin formula} for the conditionally 
negative definite function $\sqrt{\xi^2+m^2}-m$, which has an integral
representaion with a $\sigma$-finite measure $n(dy)$ 
on ${\bf R}^d \setminus \{0\}$, called {\it L\'evy measure}, which satisfies  
$\int_{|y|>0}\frac{|y|^2}{1+|y|^2} n(dy) < \infty$:
\begin{eqnarray}
 \sqrt{\xi^2+m^2} -m 
\!\!&\!=\!&\!\! -\int_{\{|y|>0\}}[e^{iy\cdot \xi} -1-iy\cdot\xi I_{\{|y|<1\}}]n(dy)
   \\
\!\!&\!=\!&\!\! -\lim_{r\rightarrow 0+} \int_{|y|\geq r}[e^{iy\cdot \xi}-1]n(dy) 
\equiv -\hbox{\rm p.v.} \int_{|y|>0}[e^{iy\cdot \xi}-1]n(dy)\,. 
\qquad \nonumber
\end{eqnarray}
Here $I_{\{|y|<1\}}$ is the indicator function of the set $\{|y|<1\}$ in
${\bf R}^d$, i.e. $I_{\{|y|<1\}}(z)=1$, if $|z|<1$, and $=0$, if $|z|\geq 1$.
 Though the L\'evy measure $n(dy)$ is $m$-dependent, we will suppress 
explicit indication of $m$-dependence so as to make the notation simpler.
$n(dy)$ has density so that $n(dy) = n(y)dy$. The density function $n(y)$ 
is given by
\begin{equation} 
n(y) =\left\{
        \begin{array}{rl}
        2\big(\frac{m}{2\pi}\big)^{(d+1)/2}\,
        \frac{K_{(d+1)/2}(m|y|)}{|y|^{(d+1)/2}}\,, & \qquad m>0,\\
        \frac{\Gamma\big(\frac{d+1}{2}\big)}{\pi^{(d+1)/2}} 
        \frac{1}{|y|^{d+1}}\,,                        &\qquad m=0,
       \end{array}\right.
\end{equation}
where $\Gamma(\tau)$ is the gamma function, and 
$K_{\nu}(\tau)$ the modified Bessel function of the third kind of order
$\nu$, which satisfies 
$0< K_{\nu}(\tau) \leq C[\tau^{-\nu}\vee \tau^{-1/2}]e^{-\tau}, \, \tau >0$
with a constant $C>0$.

To get (3.2) recall (e.g [ITa-86, Eq.(4.2), p.244]) the operator
$e^{-t[\sqrt{-\Delta+m^2}-m]}$ has integral kernel $k_0(x-y,t)$,
where
\begin{equation}
k_0(y,t) =\left\{
            \begin{array}{rl}
            2\big(\frac{m}{2\pi}\big)^{(d+1)/2}\,
            \frac{t e^{mt} K_{(d+1)/2}(m(|y|^2+t^2)^{1/2})}
                         {(|y|^2+t^2)^{(d+1)/4}}\,, &\qquad m>0,\\
            \frac{\Gamma\big(\frac{d+1}{2}\big)}{\pi^{(d+1)/2}}
            \frac{t}{(|y|^2+t^2)^{(d+1)/2}}\,, &\qquad m=0,
            \end{array}\right.
\end{equation}
und use that fact (e.g. [IkW1-62, Example 1]) that
$\frac{k_0(y,t)}{t} dy$ converges to $n(dy) = n(y)dy$, 
as measures on ${\bf R}^d$, as $t \rightarrow0+\,$.
Note that the both the expressions on the right-hand side of
$n(y)$ in (3.2) and $k_0(y,t)$ in (3.3) are continuouly connected as 
$m\rightarrow 0+$, because 
$K_{\nu}(\tau) = \frac{\Gamma(\nu)}{2}\big(\frac{2}{\tau}\big)^{\nu}(1+o(1))$
as $\tau \rightarrow 0+$.

We shall denote by $H_0 \equiv \sqrt{-\Delta+m^2}$ not only the linear 
operator of $L^2({\bf R}^d)$ into itself with domain $H^1({\bf R}^d)$
but also the linear map ${\cal F}^{-1}\sqrt{\xi^2+m^2}{\cal F}$ of
${\cal S}'({\bf R}^d)$ into itself as well as of the Sobolev space 
$H^s({\bf R}^d)$ into $H^{s+1}({\bf R}^d)$, 
where ${\cal F}$ and ${\cal F}^{-1}$ 
stand for the Fourier and inverse Fourier transforms.
Now for $f \in {\cal S}({\bf R}^d)$, put $\hat{f} = {\cal F}f$. The inverse 
Fourier transform of $\hat{f}(\xi)$ multiplied by (3.1) becomes
\begin{eqnarray}
(H_0f)(x) \!\!&\!\equiv\!&\!\! (\sqrt{-\Delta+m^2}f)(x) \nonumber\\
\!\!&\!=\!&\!\! mf(x) -\int_{\{|y|>0\}}
     [f(x+y)-f(x)-I_{\{|y|<1\}}\,y\cdot\nabla_xf(x)]n(dy).
\end{eqnarray}

Now to treat $H_A^{(1)}$, 
let $f \in C_0^{\infty}({\bf R}^d)$ and consider, for each fixed $x$,  
the function 
$$
f_x :\,\, y \, \mapsto \, e^{i(x-y)\cdot A\big(\frac{x+y}{2}\big)}f(y),
$$ 
which also belongs to $C_0^{\infty}({\bf R}^d)$. 
Replace $f$ in (3.4) by $f_x$, then we get 
\begin{eqnarray}
&&\!\!\!\!(H_0 f_x)(x) \\
 \!\!&\!=\!&\!\! 
mf(x) -\int_{\{|y|>0\}}
          [e^{-iy\cdot A(x+\frac{y}{2})}f(x+y)-f(x)-I_{\{|y|<1\}}\,
           y\cdot(\nabla_x-iA(x))f(x)]n(dy). \nonumber
\end{eqnarray}
On the other hand, notice that the left-hand side of (3.5) is written 
by use of Fourier transform
\begin{eqnarray*}
(H_0 f_x)(x)&=& \frac1{(2\pi)^d}\hbox{\rm Os--}\!
 \int\int_{{\bf R}^d\times{\bf R}^d} e^{i(x-y)\cdot\xi}\sqrt{\xi^2+m^2}
                \,f_x(y)dyd\xi \nonumber\\
&=&\frac1{(2\pi)^d}\hbox{\rm Os--}\!
 \int\int_{{\bf R}^d\times{\bf R}^d} 
 e^{i(x-y)\cdot\big(\xi +A(\frac{x+y}{2})\big)}\sqrt{\xi^2+m^2}\,f(y)dyd\xi\,, 
\end{eqnarray*}
the last member of which is nothing but the second oscillatory expression 
(2.4) for $(H_A^{(1)}f)(x)$. Thus we have obtained the identity
\begin{equation}
 (H_0 f_x)(x) = (H_A^{(1)}f)(x)
\end{equation}
for all $f \in C_0^{\infty}({\bf R}^d)$.  
This may be paraphrased ``{\sl apply $H_A^{(1)}$ to $f$} 
amounts to be the same thing as {\sl apply $H_0$ to $f_x$}". 
Thus we are lead to a new definition of $H_A^{(1)}$:
\begin{eqnarray}
([H^{(1)}_A-m]f)(x) 
&\!\!\!:=\!\!\!\!& 
  -\int_{|y|>0} \big[e^{-iy\cdot A(x+\frac{y}{2})} f(x+y) -f(x) 
          \nonumber\\
&&\qquad\qquad\qquad\,
-I_{\{|y|<1\}} y\!\cdot\! (\nabla_x-iA(x))\,f(x)\big]n(dy) \nonumber\\
&\!\!\!{}=\!\!\!\!& -\lim_{r\rightarrow 0+} \int_{|y|\geq r} 
   \big[e^{-iy\cdot A(x+\frac{y}{2})} f(x+y) -f(x)\big]\,n(dy)\nonumber\\
&\!\!\!{}\equiv\!\!\!\!& -\, \hbox{\rm p.v.}\!\int_{|y|>0} 
              \big[e^{-iy\cdot A(x+\frac{y}{2})} f(x+y) -f(x)\big]\,n(dy). 
\end{eqnarray}
This expression makes sense for $A(x)$ being more general functions 
than $C^{\infty}$. In fact, it can be shown with the Calderon--Zygmund theorem 
that the singular integral on the right-hand side of (3.7) exists
 pointwise in a.e. $x$ as well as in the $L^2$ norm, if
$A \in L^{2+\delta}_{\rm loc}({\bf R}^d;\, {\bf R}^d)$ for 
some $\delta >0$ such that 
$\int_{0<|y|<1} |A(x+\frac{y}{2})-A(x)||y|^{-d}dy$ is $L^{2}_{\rm loc}$.
Note that (3.7) reduces itself to (3.4) if $A(x) \equiv 0$.

For $H^{(2)}_A$, we can do it in the same way. Indeed, 
take $f \in C_0^{\infty}({\bf R}^d)$ to consider, 
for $x$ fixed, the function 
$$f_x :\,\,  y \, \mapsto \, 
e^{i(x-y)\cdot \int_0^1A\big((1-\theta)x+\theta y\big)d\theta}f(y),
$$ 
which belongs to $C_0^{\infty}({\bf R}^d)$. Replacing $f$ in (3.4) by $f_x$
we obtain the relation $(H_0 f_x)(x) = (H_A^{(2)}f)(x)$, namely, 
a new definition for $H^{(2)}_A$, as follows: 
\begin{eqnarray}
([H^{(2)}_A-m]f)(x) 
\!\!&\!:=\!&\!\! -\int_{|y|>0} 
  \big[e^{-iy\cdot \int_0^1 A(x+ \theta y)d\theta} f(x+y) -f(x) \nonumber\\
&&\qquad\qquad\qquad\,
   -I_{\{|y|<1\}} y\!\cdot\! (\nabla_x-iA(x))\,f(x)\big]n(dy) \nonumber\\
\!\!&\!:=\!&\!\! -\lim_{r\rightarrow 0+} \int_{|y|\geq r} 
\big[e^{-iy\cdot \int_0^1 A(x+ \theta y)d\theta} f(x+y) -f(x)\big]\,n(dy)
\nonumber\\
\!\!&\!{}\equiv\!&\!\! -\, \hbox{\rm p.v.}\!\int_{|y|>0} 
  \big[e^{-iy\cdot \int_0^1 A(x+ \theta y)d\theta} f(x+y) -f(x)\big]\,n(dy).
\qquad 
\end{eqnarray}

Now we are in a position to consider the most general definition for
the first two magnetic relativistic Schr\"odinger operators
$H^{(1)} = H_A^{(1)} +V$ and $H^{(2)} = H_A^{(2)} +V$ 
with both general vector and scalar potentials $A(x)$ and $V(x)$.
Assume that 
\begin{equation}%
 A \in  L^{1+\delta }_{\rm loc}({\bf R}^{d}) \,\, 
\hbox{for\, some} \,\,\delta >0 \,\, \hbox{and} \,\, 
V  \in  L^{1}_{\rm loc}({\bf R}^{d}), \,\, V(x)\geq 0 \,\,\, \hbox{a.e.}\,, 
\quad 
\end{equation}
or 
\begin{equation}%
A \in  L^{2+\delta}_{\rm loc}({\bf R}^{d}) \,\, 
\hbox{for\, some} \,\, \delta >0  \,\, \hbox{and} \,\, 
V  \in  L^{2}_{\rm loc}({\bf R}^{d}), \,\, V(x)\geq 0 \,\,\, \hbox{a.e.}
\end{equation}

Then, first, if $A$ and $V$ satisfy (3.10),
we can see again with the Calderon--Zygmund theorem that 
the singular integrals on the right-hand side of (3.7) and (3.8)
exist pointwise in a.e. $x$ as well as in the $L^2$ norm.

Next, if $A$ and $V$ satisfy (3.9),
multiply (3.7) and (3.8) with $\overline{u(x)}$ and integrate them by $dx$, 
then we can reach the following quadratic forms 
$h^{(1)}$ and $h^{(2)}$, 
respectively:
\begin{eqnarray}
h^{(1)}[u] &\!\!\equiv\!\!&  h^{(1)}[u,u] := h^{(1)}_{A,V}[u,u] \nonumber \\
&\!\!=\!\!& \Big(m\|u\|^{2} + {\frac{1}{2}} \int\int_{|x-y|>0}
|e^{-i(x-y)\cdot A(\frac{1}{2}(x+y))}u(x)-u(y)|^{2}n(x-y)dxdy\Big) \nonumber\\ 
&&\qquad\qquad\qquad 
+ \int V (x)|u(x)|^{2}dx =: h_A^{(1)}[u] + h_V^{(1)}[u]\,; \\
h^{(2)}[u] &\!\!\equiv\!\!&  h^{(2)}[u,u] := h^{(2)}_{A,V}[u,u]   \nonumber\\
&\!\!=\!\!& \Big(m\|u\|^{2} +  {\frac{1}{2}} \int\int_{|x-y|>0}
|e^{-i(x-y)\cdot \int_0^1 A((1-\theta)x+\theta y)d\theta}u(x)-u(y)|^{2}
n(x-y)dxdy\Big) \nonumber\\
&&\qquad\qquad\qquad + \int V (x)|u(x)|^{2}dx =: h_A^{(2)}[u] + h_V^{(2)}[u] 
\end{eqnarray}
with form domains 
$Q(h^{(j)}) = \{u \in  L^{2}({\bf R}^{d}); h^{(j)}[u] < \infty \},
\,\, j=1,2$. 
We can see with [Kat-76, VI, \S 2, Theorems 2.1, 2.6, pp.322--323] 
that under the assumption (3.9) for $A(x)$ and $V (x)$, 
there exist unique nonnegative
selfadjoint operators $H^{(j)} = H_A^{(j)} \dot{+} V$ 
({\it form sum}), $\, j=1,2$, such that
$h^{(j)}[u,v] = (H^{(j)}u,v)$ for $u,v \in  C^{\infty }_{0}({\bf R}^{d})$. 

\medskip
We expect that the condition (3.10) (resp. (3.9)) is minimal to assure that 
$H^{(j)}$ (resp. $h^{(j)})$ defines a 
linear operator (resp. quadratic form) in $L^{2}({\bf R}^{d})$ with domain 
(resp. form domain) including $C^{\infty }_{0}({\bf R}^{d})$, so long as 
$V (x)$ is nonnegative. 
Then we can show the following results under the assumptions (3.10) and (3.9). 

\medskip
\begin{thm} {\rm ([ITs2-93])} (i) 
If $A(x)$ and $V (x)$ satisfy (3.9), for each $j=1,2$, $h^{(j)}$ is a 
closed form with form domain $Q(h^{(j)})$ including 
$C^{\infty}_{0}({\bf R}^{d})$ as a form core, so that the 
minimal symmetric form $h^{(j)}_{\min}$ defined as the form closure of 
$h^{(j)}| C^{\infty }_{0}({\bf R}^{d})\times C^{\infty }_{0}({\bf R}^{d})$ 
coincides with $h^{(j)}$. Therefore there exists a unique 
selfadjoint operator $H^{(j)}= H_A^{(j)} \dot{+} V$ (form sum) 
with domain $D(H^{(j)})$ corresponding to the form $h^{(j)}$ 
such that $h^{(j)}[u,v] = (H^{(j)}u,v)$ for $u \in  D(H^{(j)})$, 
$v \in  Q(h^{(j)})$. 

(ii) If $A(x)$ and $V (x)$ satisfy (3.10), for each $j=1,2$, 
$H^{(j)}=H_A^{(j)}+V $ (operator sum) is essentially selfadjoint 
on $C^{\infty }_{0}({\bf R}^{d})$ and the closure of $H^{(j)}$, denoted by
the same $H^{(j)}$ again, is bounded from below by $m$.
\end{thm} 

The proof of statements in Theorem 3.1 for $H^{(1)}$ and $h^{(1)}$ were first 
given under less general assumption in [I3-89], [I4-93] and [ITs1-92] and 
completed as in the present form in [ITs2-93], while 
that for $H^{(2)}$ and $h^{(2)}$ given in [IfMP1-07, 2-08, 3-10] for vector 
potentials $A(x)$ which are $C^{\infty}$ functions of polynomial growth as 
$|x| \rightarrow \infty$.
 
Thus we are led to more general definitions, 
not only for the two magnetic relativistic Schr\"odinger operators 
$H^{(1)}_A$ and  $H^{(2)}_A$ than the ones in Definitions 2.3 and 2.4, 
but also for the two general relativistic Schr\"odinger operatows 
$H^{(1)}$ and  $H^{(2)}$ with both vector and scalar potentials $A(x)$ 
and $V(x)$.

\begin{dfn} 
{\rm  If $A(x)$ and $V(x)$ satisfy (3.10), for $j=1,2$, 
$H^{(j)}$ is defined as the closure of the operator sum of the integral 
operator $H_A^{(j)}$ in (3.7), (3.8) and the potential $V(x)$.
}
\end{dfn}

\begin{dfn} 
{\rm  If $A(x)$ and $V(x)$ satisfy (3.9), for $j=1,2$, 
$H^{(j)}$ is defined as the selfadjoint operator 
$H^{(j)} = H_A^{(j)} \dot{+} V$ associated with the closed 
form $h^{(j)}$ which is the sum of the two closed forms $h_A^{(j)}$ 
and $h_V^{(j)}$ as in (3.11) and (3.12).
}
\end{dfn}

Next, we come to $H_A^{(3)}$. 
If $0<\delta <1$, condition 
``$A \in L^{1+\delta}_{{\rm loc}}({\bf R}^d; {\bf R}^d)$" in (3.9) for
$A$ is slightly more general than 
condition  ``$A \in L^2_{{\rm loc}}({\bf R}^d; {\bf R}^d)$"
used to give the definition for $H_A^{(3)}$ in Definition 2.5. 
As Theorem 3.1 (ii) says that when 
$A \in L^{1+\delta}_{{\rm loc}}({\bf R}^d; {\bf R}^d)$,
$(-i\nabla-A(x))^2+m^2$ can define a nonnengative selfadjoint operator in $L^2({\bf R}^d)$, 
so we are led to the following more general definition than the one in Definition 2.5.

\begin{dfn} 
{\rm If $A(x)$ and $V(x)$ satisfy (3.9), 
$H^{(3)}$ is defined in $L^2({\bf R}^d)$ as the form sum of the square root of
the nonnegative selfadjoint operator $(-i\nabla-A(x))^2+m^2$ and $V$:
\begin{equation}%
H^{(3)} := H^{(3)}_A\,\, \dot{+}\,\, V 
         := \sqrt{(-i\nabla-A(x))^2+m^2}\,\, \dot{+}\,\, V
\end{equation}
}
\end{dfn}

\medskip
Thus, with Definitions 3.2, 3.3 and 3.4, we are now given more  general definition 
of the three relativistic Schr\"odinger operators  
$H^{(1)}$, $H^{(2)}$, $H^{(3)}$ concerned, 
corresponding to the classical relativistic symbol (1.1) with both 
vector and scalar potentials $A(x)$ and $V(x)$.

\medskip
It is appropriate here to refer, for comparison, to the corresponding results 
for the nonrelativistic magnetic Schr\"{o}dinger operator 
$H^{NR} := H_A^{NR} + V := \frac12(-i\nabla -A(x))^{2}+V (x)$. 
In fact, as alredy mentioned in Section 2.1, one can realize $H^{NR}$ as a
selfadjoint operator defined through the quadratic form  with form domain 
including $C^{\infty }_{0}({\bf R}^{d})$ as a form core when
\begin{equation}%
 A \in  L^{2}_{\rm loc}({\bf R}^{d}) \,  \hbox{and} 
\, \, V  \in  L^{1}_{\rm loc}({\bf R}^{d}), 
 \, \, V (x)\geq 0 
 \, \, \hbox{a.e.}, 
\end{equation}
which was proved by Kato and Simon (see [CFKS-87, pp.8--10]). This was  
also proved by Leinfelder-Simader [LeSi-81], who further gave a definitive 
result that it is  essentially selfadjoint on $C^{\infty }_{0}({\bf R}^{d})$ 
when 
\begin{equation}%
 A \in  L^{4}_{\rm loc}({\bf R}^{d}), \,\,
\hbox{div} A \in  L^{2}_{\rm loc}({\bf R}^{d})\,\,   
\hbox{and} \, \, V  \in  L^{2}_{\rm loc}({\bf R}^{d}),  
\, \, V (x)\geq 0  \,\, \hbox{a.e.}  
\end{equation}

The proof of Theorem 3.1 will be carried out by mimicking the arguments used 
by Leinfelder-Simader [LeSi-81]. First the statement (i) is proved. Then 
the idea of proof of the statement (ii) consists in showing that, 
when $A(x)$ and $V (x)$ satisfy (3.10), $C^{\infty }_{0}({\bf R}^{d})$ is  also 
an operator core of the selfadjoint operator $H^{(1)}$ obtained through 
the form $h^{(1)}$ in the statement 
(i). We refer the details of the proof to Ichinose-Tsuchida [ITs2-93]. 


\bigskip\noindent
{\sl Remarks.}  1$^o$. For the scalar potential 
$V \in L^{2}_{\rm loc}({\bf R}^{d})$ having negative part: 
$V (x) = V _{+}(x) - V _{-}(x)$ with $V _{\pm }(x)\geq 0$ and 
$V _{+}(x)V _{-}(x)=0$ a.e., it will be possible to show the theorem, 
if $V _{-}(x)$ is small in a certain sense 
[i.e. relatively bounded / relatively form-bounded  with respect to 
$\sqrt{-\Delta+m^2}$ or $H^{(j)}_A \, (j=1,2)$ with relative bound
less than $1$],  but we content ourselves with such $V $ as in (3.9) and (3.10). 
The main point of Theorem 3.1 is in treating the Hamiltonian with  
vector potential $A(x)$ as general as possible. 

\smallskip\noindent
2$^o$. Nagase--Umeda [NaU1-90] proved essential selfadjointness of the Weyl
pseudo-differential operator $H^{(1)}_{A}$ in (2.3). 

\smallskip\noindent
3$^o$. When $A(x)$ is in $L^{2+\delta }_{\rm loc}({\bf R}^{d}; {\bf R}^{d})$ and 
\begin{equation}%
\int_{0<|y|<1}|y\cdot (A(x+y/2)-A(x))||y|^{-d}dy \, \, \, \hbox{is\,\,in} \, 
L^{2}_{\rm loc}({\bf R}^{d}),
\end{equation}
it can be shown [ITs1-92] (cf. [I3-89]) that {\sl Kato's inequality} holds for 
$H^{(1)}_{A}$ in (3.7): If $u \in  
L^{2}({\bf R}^{d})$ with $H^{(1)}_{A} u \in  L^{1}_{\rm loc}({\bf R}^{d})$, then
the distributional inequality 
\begin{equation}
     \hbox{Re}((\hbox{sgn}\, u)[{H}^{(1)}_{A}-m]u) \geq [\sqrt{-\Delta +m^2}-m]\,|u| 
\end{equation}
holds, where (sgn$\, u)(x) = \overline{u(x)}/|u(x)|$ for 
$u(x)\neq 0\,;\, = 0$ for $u(x)=0$. In particular, if $A(x)$ is 
H\"{o}lder-continuous, then $A(x)$ satisfies condition (3.16). 
To show (3.17), one has to use the expression $(3.7)$ for $H^{(1)}_Af$ instead of (2.3).
For the detail see [Its1-92, Theorem 3.1] (cf. [I3-89, Theorems 4.1, 5.1]).

In the same way, {\sl Kato's inequality} also for $H^{(2)}_{A}$ will be shown 
with the expression (3.8) instead of (2.8): 
If $u \in  L^{2}({\bf R}^{d})$ with $H^{(2)}_{A} u \in  L^{1}_{\rm loc}({\bf R}^{d})$, 
then the distributional inequality 
\begin{equation}
     \hbox{Re}((\hbox{sgn}\, u)[{H}^{(2)}_{A}-m]u) \geq [\sqrt{-\Delta +m^2}-m]\,|u|
\end{equation}
holds. Also for ${H}^{(3)}_{A}$, we expect to have a distributional inequality 
like 
\begin{equation}
\hbox{Re}((\hbox{sgn}\, u)[{H}^{(3)}_{A}-m]u) \geq  [\sqrt{-\Delta +m^2}-m]\,|u|.
\end{equation}
However, the problem will be open. Although it can be shown [HILo2-12] (cf. [HILo1-12]) 
that inequality ({\sl Diamagnetic inequality})
\begin{equation}
 (f,e^{-t[H^{(3)}_A-m]}f) \leq (|f|, e^{-t[\sqrt{-\Delta +m^2}-m]}|f|)
\end{equation} 
holds for all $f \in L^2({\bf R}^d)$, which (cf. [S1-77], [HeScUh-77]) is equivalent 
to an abstract version of 
``Kato's inequality" for $H^{(3)}_A$, the distributional version (3.20) is a stronger
assertion. Here and throughout this article, $(\,\cdot\,, \,\cdot\,)$ in (3.20) is the 
{\sl physicist's inner product} $(f,g)$ of the Hilbert space $L^2({\bf R}^d)$, 
which is anti-linear in $f$ and linear in $g$.

\medskip
Finally, we are going to see the three magnetic relativistic 
Schr\"odinger operators $H^{(1)}_A$, $H^{(2)}_A$ and $H^{(3)}_A$ are 
bounded from below by the {\it same lower bound}, as in the following theorem. 
\begin{thm}%
\begin{equation}
 H^{(j)}_A \geq m\,, \qquad j= 1, 2, 3.
\end{equation}
\end{thm}

{\it Proof}. First, it is trivial for $H^{(3)}_A$, 
as also seen from (3.12), for instance.
Next to see for $H^{(1)}_A$, take $u \in C_0^{\infty}({\bf R}^d)$ in 
Kato's inequality (3.17) above. Multiply both sides by $|u(x)|$
and integrate them in $x$, then we have 
$$
 (u,[{H}^{(1)}_{A}-m]u) \geq (|u|,[\sqrt{-\Delta +m^2}-m]\,|u|)\,,
$$
where we note that $|u(x)|$ is in the Sobolev space $H^1({\bf R}^d)$, so that
the right-hand side above exists finite and nonnegative. 
So the assertion follows. In the same way it will be shown for $H^{(2)}_A$.
\qed

\medskip\noindent
{\sl Remark}. In the above proof, the sharp lower bound (3.21) 
for $H^{(1)}_A$ and $H^{(2)}_A$ has been obtained with their integral 
operator expressions (3.7) and (3.8). It does not seem to be obtained by
 pseudo-differential calculus from their expressions (2.2)/(2.3) 
and (2.8)/(2.9), but instead then probably only a bound such as 
$H_A^{(j)} \geq m-\delta$ for some $\delta >0$ 
(cf. [Ho-85, Section 18.1]).


\subsection{Selfadjointness with negative scalar potentials} 

We have seen above that our relativistic Schr\"odinger operators with 
{\it nonnegative} scalar potentials  
assume analogous aspects on selfadjointness problem
with the nonrelativistic Schr\"{o}dinger operator. 
In this subsection we shall observe that, as for {\it negative} 
scalar potentials unbounded at infinity, the former makes a remarkable 
contrast with the latter, bearing an aspect closer to the Dirac operator 
(cf. Chernoff [Ch-77]), though the relativistic Schr\"{o}dinger equation 
$i\frac{\partial}{\partial t}\varphi (x,t) = [H_{0}-m]\varphi(x,t)$ 
has not the finite propagation property. 

For comparison, first we refer to the result for the nonrelativistic 
Schr\"{o}dinger operator $-\Delta + V(x) + W(x)$ in $L^2({\bf R}^d)$ by
Faris--Lavine [FaLa-74] (or [RS-75, Theorem X.38, p.198]): 
Assume that the real-valued scalar potentials $V$ and $W$ obey
the following conditions:
Let $V$ be in $L^p_{\rm loc}({\bf R}^d)$ for some 
$p \geq (d/2)\vee 2 \,\,(p>2\,\, \hbox{\rm when}\,\, d=4)$, and 
let $W$ be in $L^2_{\rm loc}({\bf R})$ and satisfy
$W(x) \geq -c_1|x|^2 -c_2$ for some constants $c_1,\,\, c_2$, 
then $-\Delta + V(x) + W(x)$ is essentially selfadjoint on
$C_0^{\infty}({\bf R}^d)$.

Now we consider the magnetic relativistic Schr\"{o}dinger operator 
$H^{(1)} = H^{(1)}_{A} + V + W$, assuming the following conditions: 
Let $A$ be in $L^{2+\delta }_{{\rm loc}}({\bf R}^{d};\,{\bf R}^{d})$ 
for some $\delta >0$ and satisfies (3.16), and $V $  
in $L^{2}_{{\rm loc}}({\bf R}^{d})$ which is relatively bounded
with respect to $H_{0} \equiv \sqrt{-\Delta + m^2}$ with 
relative bound $< 1$. Let $W$ be in $L^{2}_{{\rm loc}}({\bf R}^{d})$. 
Assume further with constants $a\geq 0,\,\, 0<b\leq 1\,\,\, c\geq 0\,$ that 
\begin{equation}%
A(x)\,\,\hbox{\rm is bounded by}\,\,
\hbox{\rm a polynomial of}\,\, |x|\,\, \hbox{\rm and}\,\, 
W(x) \geq  -c \exp[a| x|^{1-b}]\,,
\end{equation}
\noindent
\hbox{\rm or}
\begin{equation}
\quad\quad A(x)\,\,\hbox{\rm is bounded and}\,\, W(x) \geq  -c e^{a|x|}\,. 
\end{equation}

\medskip\noindent
Note that condition (3.22) or (3.23) allows $W(x)$ to decrease 
exponentially at infinity with respect to $|x| ^{1-b}$ or $| x| $. 

Then we have the following result in [Iw-94] and [IIw-95]  
for the magnetic relativistic Schr\"{o}dinger operator 
$H^{(1)} = H^{(1)}_{A} + V + W$. The former work used, for the operator $H^{(1)}_A$,
the pseudo-differential operator expression in Definition 2.3 with (2.3), while
the latter the integral operator expression connected with L\'evy process 
in Definition 3.2 with (3.7). The latter result is sharper than the former.   

\medskip\noindent
\begin{thm} {\rm ([IIw-95])}
 With the above assumption on $A(x)$, $V (x)$ and $W(x)$, 
$H^{(1)} :=  H^{(1)}_{A} + V  + W$ is essentially selfadjoint on 
$C^{\infty }_{0}({\bf R}^{d})$. 
\end{thm}

\medskip\noindent
{\sl Example}. For 0 $\leq  Z < (d-2)/2$, $d\geq 3$, the operator 
$H^{(1)}_{A} - Z/|x|  + W(x)$ with H\"{o}lder-continuous $A(x)$ 
 and locally square-integrable $W(x)$ satisfying (3.22) or (3.23) 
is essentially selfadjoint on $C^{\infty}_{0}({\bf R}^{d})$.

\medskip
{\it Proof of Theorem 3.6}. (Sketch) First, using that $V $ is 
$H_{0}$-bounded with relative bound $< 1$, we show that $H^{(1)}_{A} + V $ is a 
symmetric operator 
with domain $C^{\infty }_{0}({\bf R}^{d})$ bounded from below. 

Next, we use Kato's inequality (3.17) for $H^{(1)}_{A}$ to show 
that if $W(x) \geq 0$ is in $L^{2}_{{\rm loc}}({\bf R}^{d})$, 
the range of $R  + H^{(1)}_{A} + V  + W$ is dense for 
$R >0$ sufficiently large, i.e., that $H^{(1)}_{A} + V  + W$ is 
essentially selfadjoint on $C^{\infty }_{0}({\bf R}^{d})$.

Then we apply the arguments used by Faris-Lavine [FaLa-74]. The following 
lemma will be needed on the commutators of $H^{(1)}_{A}$ with the square 
roots of the exponential functions in (3.22) and (3.23) bounding 
$W(x)$ from below. 

\medskip\noindent
\begin{lemma}
 If $A(x)$ and $V(x)$ satisfy (3.22) (resp. (3.23)), then, for $\psi (x) =$ 
$\exp[(a/2)(1+x^{2})^{(1-b)/2}]$ with $a\geq 0$ and $0<b\leq 1$ 
(resp. $\psi (x) =$ $\exp[(a/2)(1+x^{2})^{1/2}]$ with 0 $\leq  a/2 < m)$, 
there exists a constant $C \geq $ 0 such that 
$$\|[H^{(1)}_{A} , \psi ]u\|  \leq  C\| \psi u\| , \qquad u \in  
C^{\infty }_{0}({\bf R}^{d}),
$$
where  $[H^{(1)}_{A} , \psi] := H^{(1)}_{A}\psi - \psi H^{(1)}_{A}$.
\end{lemma}

The proof of Lemma 3.7 is omitted and referred to [IIw-95].

\medskip
We continue the proof of Theorem 3.6.  Choose a nonnegative constant $K$ such 
that $K \geq  2C$ and $V(x) +$ 
$K\psi (x)^{2} \geq $ 0 a.e., where $C$ and $\psi $ are the same constant and 
the same function as in Lemma 3.7. Let $N = H^{(1)} + 3K\psi ^{2} = 
H^{(1)}_{A} + V  + W + 3K\psi ^{2}$, which is, as has been 
seen above, essentially selfadjoint on $C^{\infty }_{0}({\bf R}^{d})$. 
The closure of $N$ is denoted by the same $N$. 
Then $N$ satisfies $N \geq  2K\psi ^{2}$ and has 
$C^{\infty }_{0}({\bf R}^{d})$ as its 
operator core. To prove essential selfadjointness of $H^{(1)}$ we have only 
to show that $\|H^{(1)}u\|  \leq  \|Nu\| $ for 
$u \in  C^{\infty }_{0}({\bf R}^{d})$ and that 
$\pm i[H,N] \leq  3CN$ 
as the quadratic forms. This can be shown with the aid of the above lemma. 

Finally we end the proof with the following note. When $A(x)$ and 
$V(x)$ satisfy (3.23), the above lemma appears to restrict the lower bound 
function $Ce^{a| x| }$ by $0\leq a/2<m$, but it is not so. To see this, for the moment 
only here we write $H^{(1)}_{A}$ in (3.7) as $H^{(1)}_{A,m}$ so as to manifest 
its $m$-dependence. Then recall [I4-92, Theorem 2.3] that the difference 
$H^{(1)}_{A,m} - H^{(1)}_{A,m'}$ is a bounded operator for all $m$, 
$m' \geq  0$. 
Therefore, if it is shown that $H^{(1)}_{A,m} + V  + W$ is essentially 
selfadjoint for some $m\geq 0$, then it follows by the Kato--Rellich theorem 
that so is $H^{(1)}_{A,m'}+ V  + W$ for every $m'\geq 0$. 
\qed

\medskip
Results similar to Theorem 3.6 will hold for $H^{(2)}$ and $H^{(3)}$.

\section{Imaginary-time path integrals for  
magnetic relativistic Schr\"odinger operators} 

It is well-known that the solution $u(x,t)$ of the Cauchy problem for 
the heat equation 
$
\frac{d}{dt}u(x,t) = \Big[\frac{1}{2}\Delta -V(x)\Big]u(x,t) 
$
with initial data $u(x,0)=g(x)$ 
can be represented by path integral, called {\it Feynman--Kac formula}
(e.g. [RS-75, Theorem X.68, p.279], 
[Demuth--van Casteren [DvC-00], Theorem 2.5, p.61):
\begin{eqnarray}  
u(x,t) 
&\!=\!& (e^{-t[-\frac{1}{2}\Delta +V]}g)(x) 
\!=\! \int_{C_x([0,\infty)\rightarrow {\bf R}^d)} 
   \!\!e^{-\int_0^tV(B(s))ds} g(B(t)) d\mu_x(B), \quad
\end{eqnarray}
where, for each $x \in {\bf R}^d$, $\mu_x$ is the {\it Wiener measure}  on 
the space $C_x([0,\infty)\,\,\rightarrow\,{\bf R}^d)$ of the Brownian paths
 which are continuous functions $B: [0,\infty) \rightarrow {\bf R}^d$
satisfying $B(0)=x$ (See also (4.29) below). 
The stochastic process concerned is called {\it Wiener process}. 
As $-\frac{1}{2}\Delta + V$ is a nonrelativistic
Schr\"odinger operator with scalar potential 
\begin{figure}[h]
\begin{center}
\includegraphics[width=10cm,clip]{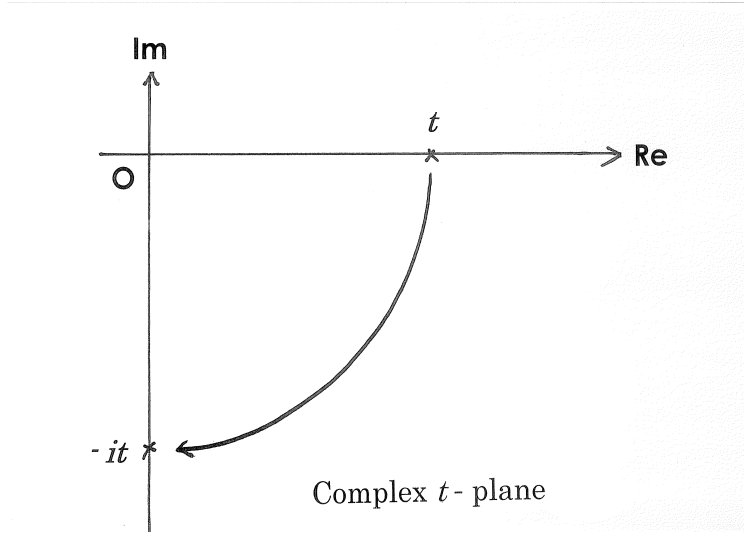}
\caption{\small{From real time $t$ to imaginary time $-it$}}
\end{center}
\end{figure}
\noindent
$V(x)$ with mass $1$, 
so the heat equation can be thought to be 
{\sl imaginary-time Schr\"odinger equation},
because it is the equation to be obtained by starting
from (real-time) Schr\"odinger equation 
$
i\frac{\partial}{\partial t} \psi(x,t) 
       = \Big[-\frac{1}{2}\Delta +V(x)\Big]\psi(x,t), 
$
next rotating it by $- 90^o$ from real time $t$ to imaginary time $-it$ 
in complex $t$-plane (see Figure 1) (cf. [I5-93, Section  4, p.23]) and then by
{\sl formally} putting $u(x,t) := \psi(x,-it)$, however, 
without seriously thinking about its meaning.

For a general nonrelativistic Schr\"odinger operator with vector and 
scalar potentials $A(x)$ and $V(x)$, 
$H^{NR}:= H_A^{NR}+ V := \frac{1}{2}(-i\nabla -A(x))^2 +V(x)\, $,  
there is also a path integral representation, called 
{\it Feynman--Kac--It\^o formula} (e.g. Simon [S2-79/05]),
for the solution $u(x,t) = (e^{-tH^{NR}}\!g)(x)$ 
of the Cauchy problem for the imaginary-time nonrelativistic Schr\"odinger 
equation, i.e. corresponding heat equation  
$\frac{d}{dt}u(x,t) = -H^{NR} u(x,t)$ with initial data $u(x,0)= g(x)$:
\begin{eqnarray}  
&&(e^{-tH^{NR}}\!g)(x) \nonumber\\ 
&\!=\!& \int_{C_x([0,\infty)\rightarrow {\bf R}^d)} 
   \!\!e^{-[i\int_0^t A(B(s)) dB(s) 
      +\frac{i}{2} \int_0^t \hbox{\footnotesize div} A(B(s)) ds 
      +\int_0^tV(B(s))ds]} \!g(B(t)) d\mu_x(B)\nonumber\\
&\!\equiv\!& \int_{C_x([0,\infty)\rightarrow {\bf R}^d)} \,
   \!e^{-[i\int_0^{t} A(B(s))\circ dB(s)+\int_0^tV(B(s))ds]}g(B(t))\, d\mu_x(B),
\end{eqnarray}
where $\int_0^t A(B(s)) dB(s)$ is {\it It\^o's integral}
and $\int_0^{t} A(B(s))\circ dB(s)$ the {\it Stratonovich integral}.
In other words, we can say that these formulas (4.1) and (4.2) 
are representing the nonrelativistic Schr\"odinger semigroups 
$e^{-t[-\frac{1}{2}\Delta +V]}$ and $e^{-tH^{NR}}$.  

In this section, we consider the same problem for the three
magnetic relativistic relativistic Schr\"odinger operators 
$H^{(1)}$,  $H^{(2)}$ and $H^{(3)}$. In Section 4.1 
we give path integral represetations for their respective semigroups, 
and in Section 4.2 we discuss how these formulas 
are be able to be deduced through some heuristic consideration. 

\subsection{Feynman--Kac--It\^o type formulas for magnetic relativistic 
Schr\"odinger operators}

Let $H$ be one of the magnetic relativistic Schr\"odinger operators  
$H^{(1)}, \, H^{(2)}, \, H^{(3)}$ in Definitions 2.1, 2.2, 2.3 or
Defintions 3.2, 3.3, 3.4. 
In the same way as in the nonrelativistic 
case, rotating (real-time) relativistic Schr\"odinger equation
$i\frac{\partial}{\partial t} \psi(x,t) = [H-m]\psi(x,t)$ by 
$- 90^o$ from real time $t$ to imaginary time $-it$ in complex $t$-plane
(cf. [I5, Section 4, p.23]), we arrive at the {\sl imaginary-time 
relativistic Schr\"odinger equation}, i.e. the corresponding 
``heat equation" for $H-m$  [formally putting $u(x,t) := \psi(x,-it)$]:
\begin{equation}
\left\{
 \begin{array}{rl}
 \displaystyle{\frac{\partial}{\partial t}} u(x,t) &= -[H-m]u(x,t), 
                                 \quad t>0,\\
     u(x,0) &= g(x), \quad\qquad\qquad\quad x \in {\bf R}^d.
 \end{array}\right.
\end{equation}
The semigroup $u(x,t) = (e^{-t[H-m]}g)(x)$ gives the solution of 
this Cauchy problem as well. We want to deal with 
 path integral representation for each $e^{-t[H^{(j)}-m]}g\, (j=1,2,3)$.
The relevant path integral is connected with the 
{\it L\'evy process} (e.g. Ikeda--Watanabe [IkW2-81/89], Sato [Sa2-99], 
Applebaum [Ap-04/09]) 
on the space 
$D_x := D_x([0,\infty)\rightarrow {\bf R}^d)$, with each $x \in {\bf R}^d$, 
of the {\it ``c\`adlag} paths", i.e. right-continuous paths
$X: \![0,\infty)\! \ni s \mapsto \! {\bf R}^d$ having left-hand limits and 
with $X(0)\!=\!x$. 
The associated path space measure is a probability measure
 $\lambda_x$, for each $x \in {\bf R}^d$, 
on $D_x([0,\infty)\rightarrow {\bf R}^d)$  whose characteristic function 
is given by 
\begin{equation}%
e^{-t[\sqrt{\xi^2+m^2}-m]} 
 = \int_{D_x([0,\infty)\rightarrow {\bf R}^d)} 
  e^{i (X(t)-x)\cdot\xi}d\lambda_x(X), \quad t\geq 0, 
\quad \xi \in {\bf R}^d.
\end{equation}
This path integral formula with measure $\lambda_x$ was effectively used 
by [CaMS-90] to get aysmptotic behavior of eigenfunctions for 
relativistic Schr\"odinger operator without vector potential.
It also, together with the Feyman--Kac formula (4.2) with Wiener measure 
$\mu_x$, was powerfully used in [ITak1-97, 2-98]
to estimate in norm the difference between the Kac transfer operator 
$e^{-tV/2}e^{-tH_A}e^{-tV/2}$ and the
nonrelativistic and/or relativistic Schr\"odinger semigroup $e^{-t(H_A+V)}$
by a power of $t$, in the case that $H_A$ is a nonrelativistic magnetic 
Schr\"odinger operator $H_A^{NR}$ and/ or a free relativistic Schr\"odinger 
operator $H_0 = \sqrt{-\Delta+1}-1$ with mass $m$.

We are going to start on task of representing the semigroup 
$e^{-t[H-m]}g$ by path integral. 
Before that, let us note that when the vector potential $A(x)$ is absent,
we can represent $u(x,t)$ by a formula looking similar to  
the Feynman--Kac formula (4.1) 
for the nonrelativistic Schr\"odinger equation: 
\begin{equation}
u(x,t) =(e^{-t[\sqrt{-\Delta+m^2} +V-m]}g)(x) 
       = \int_{D_x([0,\infty)\rightarrow{\bf R}^d)}
        e^{-\int_0^t V(X(s))ds} g(X(t)) d\lambda_x(X).
\end{equation}

\bigskip
Now, when the vector potential $A(x)$ is present,
let us treat each case for $H^{(1)}$, $H^{(2)}$ and $H^{(3)}$, separately.

\medskip\noindent
(1) First consider the case for $H^{(1)}:= H^{(1)}_A + V$ 
in Definition 3.3 with condition (3.9) on $A(x)$ and $V(x)$.

To represent $e^{-t[H^{(1)}-m]}g$ by path integral, we need some further 
notations from L\'evy process.  

For each path $X$, $N_X(dsdy)$ denotes the {\it counting measure} on 
$[0,\infty)\times ({\bf R}^d\setminus \{0\})$ to count the number 
of discontinuiies of $X(\cdot)$, i.e. 
\begin{equation}
N_X((t,t']\times U) := \#\{s \in (t,t'];\, 0\neq X(s)-X(s-) \in U\}
\end{equation}
with $0 <t <t'$ and $U \subset {\bf R}^d\setminus \{0\}$ being a Borel set. 
It satisfies 
$\int_{D_x}N_X(dsdy)\, d\lambda_x(X) =dsn(dy)$. 
Put 
\begin{equation}
\widetilde{N}_X(dsdy) = N_X(dsdy) -dsn(dy),
\end{equation}
which may be thought of as a renormalization of $N_X(dsdy)$.
Then any path $X \in D_x([0,\infty)\rightarrow {\bf R}^d)$ can be
expressed with
$N_x(\cdot)$ and $\widetilde{N}_X(\cdot)$ as
\begin{eqnarray}
X(t) -x &=& \int_0^{t+}\int_{|y|\geq 1} y N_X(dsdy) 
               + \int_0^{t+}\int_{0<|y|<1} y \widetilde{N}_X(dsdy)
                         \nonumber\\
      &=& \int_0^{t+}\int_{|y|>0} y \widetilde{N}_X(dsdy).
\end{eqnarray}

Then we have the following path integral representation for 
$e^{-t[H^{(1)}-m]}g$.

\begin{thm} {\rm ([ITa-86], [I7-95])} 
Assume that $A(x)$ and $V(x)$ satisfy condition (3.9). Then
\begin{eqnarray}
(e^{-t[H^{(1)}-m]}g)(x) \!&\!\!=\!\!&\! 
 \int_{D_x([0,\infty)\rightarrow {\bf R}^d)} e^{-S^{(1)}(X; x,t)} 
  g(X(t))\,d\lambda_x (X),\\ 
S^{(1)}(X; x,t)\!&\!\!=\!\!&\! i\int_0^{t+}\int_{|y|\geq 1} 
                 A(X(s-)+\frac{y}{2})\!\cdot\! y\, N_X(dsdy) \nonumber\\
 &\!&  +i\int_0^{t+}\int_{0<|y|<1} 
            A(X(s-)+\frac{y}{2})\!\cdot\! y\, \widetilde{N}_X(dsdy)
                                                             \nonumber\\
 &\!& +i\int_0^{t} \int_{0<|y|<1}
              [A(X(s)+\frac{y}{2})-A(X(s))]\!\cdot\! y\, dsn(dy) 
                                     \!+\! \int_0^t V(X(s))ds     \nonumber\\
 \!&\!\!=\!\!&\! i\int_0^{t+}\int_{|y|>0} 
            A(X(s-)+\frac{y}{2})\!\cdot\! y\, \widetilde{N}_X(dsdy)\\
 &\!& +i\int_0^{t} \int_{|y|>0}
              [A(X(s)+\frac{y}{2})-A(X(s))]\!\cdot\! y\, dsn(dy) 
                   \!+\! \int_0^t V(X(s))ds.                       \nonumber
\end{eqnarray}
\end{thm}

\bigskip
Here it is easy to see the second equality in the expression (4.10) 
for $S^{(1)}(X; x,t)$, as well as in (4.8). We note also that, in (4.10), 
the integral in the third term of its second member is also 
written as the principal value integral:
$$
\int_0^{t} \int_{0<|y|<1}
                     [A(X(s)+\frac{y}{2})-A(X(s)]\!\cdot\! y\, dsn(dy)
= \int_0^{t} ds\, \hbox{\rm p.v.}\! \int_{0<|y|<1} 
                 A(X(s)+\frac{y}{2})\!\cdot\! y\, n(dy),
$$
and the same is valid for the second term  of its third (last) member.

\bigskip
{\it Proof of Theorem 4.1}.  We shall show first the case that 
both $A$ and $V$ are bounded and smooth, precisely, 
$A \in C_b^{\infty}({\bf R}^d; {\bf R}^d)$ 
and $V \in C_b^{\infty}({\bf R}^d; {\bf R})$, 
where, for $l$ an positive integer, $C_b^{\infty}({\bf R}^d; {\bf R}^l)$ is 
the Fr\'echet space of the ${\bf R}^l$-valued $C^{\infty}$ functions 
in ${\bf R}^d$ which together with their derivatives of all orders are bounded. 
Then we shall show the general case where they satisfy condition (3.9).
Our proof follows the spirit of the proof of [RS-75, Theorem X.68, p.279] 
and [S2-79/05]. Representing the set $\Omega$ 
of ``random variables $\omega$" by the path space $D_x([0,\infty)\rightarrow{\bf R}^d)$,
we are suppressing use of ``random variable $\omega$"
by identifying it with path $X$.

\medskip
I. The case that  
$A \in C_b^{\infty}({\bf R}^d; {\bf R}^d)$ 
and $V \in C_b^{\infty}({\bf R}^d; {\bf R})$.

Introduce a bounded operator $T(t)$ on $L^2({\bf R}^d)$ by 
\begin{eqnarray}
(T(t)g)(x) 
\!\!&\!:=\!&\!\! \displaystyle{\int_{{\bf R}^d}} k_0(x-y,t)
  e^{iA\big(\frac{x+y}{2}\big)\cdot(x-y)-V\big(\frac{x+y}{2}\big)t}g(y)dy
  \nonumber\\
\!\!&\!\,=\!&\!\! \displaystyle{\int_{{\bf R}^d}} k_0(x-y,t)
  e^{-iA\big(\frac{x+y}{2}\big)\cdot(y-x)-V\big(\frac{x+y}{2}\big)t}g(y)dy
\end{eqnarray}
where $k_0(x-y,t)$ is the integral kernel of $e^{-t[\sqrt{-\Delta+m^2}-m]}$ in (3.3),
which is nonnegative and satisfies $k_0(-x,t) = k_0(x,t)$. Then we can rewrite 
$T(t)$ as
\begin{equation}
(T(t)g)(x) = \int_{D_x} 
  e^{-iA\big(\frac{x+X(t)}{2}\big)\cdot(X(t)-x)
             -V\big(\frac{x+X(t)}{2}\big)t}g(X(t))d\lambda_x(X)\,.
\end{equation}

Do partition of $[0,t]$: 
$0=t_0<t_1< \cdots < t_n=t,\,\, t_j-t_{j-1}=t/n$, and put
\begin{equation}
S_n(x_0,\cdots,x_n)
:= i\sum_{j=1}^n A\big(\frac{x_{j-1}+x_j}{2}\big)\cdot(x_{j-1}-x_{j})
   +\sum_{j=1}^n V\big(\frac{x_{j-1}+x_j}{2}\big)\frac{t}{n},
\end{equation}
where 
$x_j=X(t_j)\, (j=0,1,2,\dots,n)$; $\,x=x_0=X(t_0)\equiv X(0)$, 
$\,y=x_n=X(t_n)\equiv X(t)$. 
Substitute these $n+1$ points $x_j= X(t_j)$ on the path $X(\cdot)$ into 
$S_n(x_0,\cdots,x_n)$ to get
\begin{eqnarray}
S_n(X) 
\!\!&\!:=\!&\!\! S_n(X(t_0),\cdots,X(t_n)) \nonumber\\
\!\!&\!\,=\!&\!\! i\sum_{j=1}^n 
 A\Big(\frac{X(t_{j-1})\!+\!X(t_j)}{2}\Big)\cdot(X(t_{j-1})\!-\!X(t_{j}))  
+\sum_{j=1}^n V\Big(\frac{X(t_{j-1})\!+\!X(t_j)}{2}\Big)\frac{t}{n}\,.
\nonumber\\
&&
\end{eqnarray}
Then we have 
\begin{eqnarray}
\big(T({t}/{n})^n g\big)(x)
\!\!&\!=\!&\!\! \overbrace{\int_{{\bf R}^d}\cdots\int_{{\bf R}^d}}
       ^{\mbox{$n$ times}}
 \prod_{j=1}^n k_0(x_{j-1}-x_j,t/n)
  e^{-S_n(x_0,\cdots,x_n)}g(x_n)dx_1\cdot \cdots \cdot dx_{n} 
                                              \qquad\nonumber\\
\!\!&\!=\!&\!\! \int_{D_x} e^{-S_n(X)}g(X(t))\, d\lambda_x(X),
\qquad x_0=x.
\end{eqnarray}

Before we continue further the proof of Theorem 4.1, we show the 
following proposition which refers to the convergence of the left-hand side 
of (4.15).

\begin{prp} 
 $T({t}/{n})^n$ converges strongly to $e^{-t[H^{(1)}-m]}$  in $L^2({\bf R}^d)$
as  $n\rightarrow \infty$.  
\end{prp}

\bigskip
{\it Proof}. Since the operators $T(t/n)^n$ are uniformly bounded, 
we have only to show that $T(t/n)^ng$ is convergent to the limit in $L^2({\bf R}^d)$ 
for $g$ in the domain $D[H^{(1)}] = H^1({\bf R}^d)$ of $H^{(1)}$. 
We have with 
$M = {\sup}_{x \in {\bf R}^d} |V(x)|$ 
\begin{eqnarray*}
&\!\!&\!\! \|T(t/n)^n g - e^{-t[H^{(1)}-m]} g \| \\
\qquad&\!=\!&\!\! \| \sum_{j=1}^n T(t/n)^{j-1}(T(t/n) 
                    - e^{-(t/n)[H^{(1)}-m]})e^{-(n-j)(t/n)[H-m]} g \|\\
\qquad&\!\leq\!&\!\! e^{Mt} \sup_{0 \leq s \leq t} n \|(T(t/n) 
                    - e^{-(t/n)[H^{(1)}-m]})e^{-s[H^{(1)}-m]} g\|, 
\end{eqnarray*}
which we can show tends to zero uniformly on each bounded $t$-interval 
in $[0, \infty)$ as $n \rightarrow \infty$. To see this, we show first
that $(d/d\tau )(T(\tau)g)$ converges to $-[H-m]g$ in $L^{2}$,
as $\tau\downarrow 0$. Indeed, 
we have by (3.4)
\begin{eqnarray*}
&&\int \big([\sqrt{-\Delta_x+m^2}-m]k_0(x-y,\tau)\big)
     e^{-iA\big(\frac{x+y}{2}\big)\cdot(y-x)-V\big(\frac{x+y}{2}\big)\tau}g(y)dy\\
\!\!&\!=\!&\!\! -\int\int_{|z|>0} [k_0(x+z-y,\tau)-k_0(x-y,\tau) 
  -I_{\{|z|<1\}}z\cdot\nabla_xk_0(x-y,\tau)]n(dz)\\
&&\qquad\qquad\qquad\qquad \times 
  e^{-iA\big(\frac{x+y}{2}\big)\cdot(y-x)-V\big(\frac{x+y}{2}\big)\tau}g(y)dy\\
\!\!&\!=\!&\!\! -\int\int_{|z|>0} \Big[k_0(x-y,\tau)
 e^{-iA\big(\frac{x+z+y}{2}\big)\cdot(z+y-x)-V\big(\frac{x+z+y}{2}\big)\tau}
g(z+y)\\
&&-\Big(k_0(x-y,\tau) -I_{\{|z|<1\}}z\cdot\nabla_x k_0(x-y,\tau)\Big)
 e^{-iA\big(\frac{x+y}{2}\big)\cdot(y-x)-V\big(\frac{x+y}{2}\big)\tau}g(y)\Big]n(dz)dy,
\end{eqnarray*}
where we have changed the variable $z-y =: -y'$ and then rewritten $y$ for $y'$ again. 
Then noting that $\nabla_x k_0(x-y,\tau) = -\nabla_y k_0(x-y,\tau)$ and 
integrating by parts in the variable $y$, we see that the above integral
converges to
$$- \int_{|z|>0}[e^{-iz\cdot A\big(x+\frac{z}{2}\big)}
g(x+z) -g(x) -I_{\{|z|<1\}}z\cdot\nabla_x g(x)] n(dz),
$$
as $\tau \rightarrow +0$, because then $k_0(x-y,\tau) \rightarrow \delta(x-y)$. 
It follows that, as $\tau \rightarrow +0$, 
\begin{eqnarray*}
&&\big(\frac{\partial}{\partial \tau} T(\tau)g)\big(x)\\
&=& -\int_{|y|>0}
 \Big[\Big(\big[\sqrt{-\Delta_x+m^2}-m\big]+V(\frac{x+y}{2}\Big)k_0(x-y,\tau)\Big]
     e^{-iA\big(\frac{x+y}{2}\big)\cdot(y-x)-V\big(\frac{x+y}{2}\big)\tau}g(y)dy
\qquad
\end{eqnarray*}
converges to 
$$- \int_{|z|>0}\big[e^{-iA\big(x+\frac{z}{2}\big)\cdot z}
g(x+z) -g(x) -I_{|z|<1}z\cdot\nabla_x g(x)\big] n(dz) -V(x)g(x)
= (-[H^{(1)}-m]g)(x).
$$ 
Thus we see that for $h \in D[H^{(1)}]$
\begin{eqnarray*}
n\|[T(t/n)-e^{-(t/n)[H^{(1)}-m]}] h\|
&=& n\Big\|\int_0^{t/n}\frac{\partial}{\partial \tau}
 \big[T(\tau)-e^{-\tau[H^{(1)}-m]}\big]h d\tau\Big\|\\ 
&=& n\Big\|\int_0^{t/n}\Big(\frac{\partial}{\partial \tau}T(\tau) +[H^{(1)}-m]
   e^{-\tau[H^{(1)}-m]}\Big)h d\tau\Big\| \\
&\leq& t\sup_{0\leq \tau \leq t/n}\Big\|
\Big(\frac{\partial}{\partial \tau}T(\tau) +[H^{(1)}-m]e^{-\tau[H^{(1)}-m]}\Big)h \Big\|  
\end{eqnarray*}
converges to zero as $n\rightarrow \infty$. Moreover, this convergence is uniform
on compact subsets in $t\geq 0$ as $n\rightarrow \infty$. 
Noting $D[H^{(1)}]$ is a Hilbert space $D[H^{(1)}]$ 
with graph norm of $H^{(1)}-m$, we see by uniform boundedness principle 
that the sequence
$\{n\big(T(t/n)-e^{-(t/n)[H^{(1)}-m]}\big)\}_{n=1}^{\infty}$ is, as a family 
of bounded operators of the Hilbert space $D[H^{(1)}]$ into $L^2({\bf R}^d)$, 
uniformly bounded for all $n$  and on every fixed compact subset in $t\geq 0$. 
Consequently, it converges to zero uniformly on compact subsets of the Hilbert 
space 
$D[H^{(1)}]$. The map $[0,t]\ni s \mapsto e^{-s[H^{(1)}-m]}\in D[H^{(1)}]$
is continuous, so that $\{ e^{-s[H^{(1)}-m]}g;\, 0\leq s \leq t\}$ is a compact subset
of $D[H^{(1)}]$. This shows Proposition 4.2.
\qed

\bigskip
We continue the proof of Theorem 4.1, I.

To see the convergence of the right-hand side of (4.15), put 
\begin{eqnarray}
 S_{n}(X) &=& S_{n1}(X) + S_{n2}(X),\\
 S_{n1}(X) &=& i\sum ^{n}_{j=1} A\Big({\frac{X(t_{j-1})+X(t_{j})}{2}} \Big) 
 \cdot (X(t_{j})-X(t_{j-1})),  \\
 S_{n2}(X) &=& \sum ^{n}_{j=1} V\Big({\frac{X(t_{j-1})+X(t_{j})}{2}}\Big) 
 (t_{j}-t_{j-1}). 
\end{eqnarray}

First, for $S_{n2}(X)$ in (4.18), it is evident that for each $X \in D_{x}$,  
 $S_{n2}(X)$ converges to $\int ^{t}_{0} V(X(s))ds$, i.e. 
the last term of the second member of (4.10), as 
$n \rightarrow  \infty $. 

Next, to see the convergence of $S_{n1}(X)$ in (4.17) to the sum of
the other three terms in the same (second) member of (4.10) which involve
$A(\cdot)$, we rewrite by 
 It\^{o}'s formula [IkW2-81/89, Chap.II, 5, Theorem 5.1] (cf. (4.8)) 
the summand in 
$S_{n1}(X)$ as 

\begin{eqnarray*}
&&\!\!\! A\Big({\frac{X(t_{j-1})+X(t_{j})}{2}}\Big) \cdot (X(t_{j})-X(t_{j-1}))\\ 
&=&\!\!\! {\displaystyle \int_{t_j-1}^{t_j+} \int_{|y|>0}} 
  \Big[A\Big(\frac{X(s-)+X(t_{j-1})+yI_{|y|\geq 1}(y)}{2}\Big)
      \cdot\big(X(s-)-X(t_{j-1})+yI_{\{|y|\geq  1\}}(y)\big)\\
&&\qquad\qquad\qquad
- A\Big(\frac{X(s-)+X(t_{j-1})}{2}\Big)\cdot\big(X(s-)-X(t_{j-1})\big)
    \Big]\,N_X(dsdy)\\
{}\!\!\!&+&\!\!\!\, {\displaystyle \int_{t_j-1}^{t_j+}\int_{|y|>0}} \Big[
  A\Big(\frac{X(s-)+X(t_{j-1})+yI_{|y|< 1}(y)}{2}\Big)
      \cdot\big(X(s-)-X(t_{j-1})+yI_{\{|y|< 1\}}(y)\big)\\
&&\qquad\qquad\qquad
- A\Big(\frac{X(s-)+X(t_{j-1})}{2}\Big)\cdot\big(X(s-)-X(t_{j-1})\big)
    \Big] \,\widetilde{N}_X(dsdy)\\
{}\!\!\!&+&\!\!\!\,{\displaystyle \int_{t_j-1}^{t_j}\int_{|y|>0}}
  \Big\{A\Big(\frac{X(s)+X(t_{j-1})+yI_{|y|< 1}(y)}{2}\Big)
    \cdot\big(X(s)-X(t_{j-1})+yI_{\{|y|< 1\}}(y)\big)\\
&&\qquad  
- A\Big(\frac{X(s)+X(t_{j-1})}{2}\Big)\cdot\big(X(s)-X(t_{j-1})\big)\\
&&\qquad\qquad -I_{\{|y|<1\}}(y)\Big[\Big(\frac12(y\cdot\nabla)A\Big)
    \Big(\frac{X(s)+X(t_{j-1}}{2}\Big) \cdot\big(X(s)-X(t_{j-1})\big)\\
&&\qquad\qquad\qquad 
 + y\cdot A\Big(\frac{X(s)+X(t_{j-1})}{2}\Big)\Big]\Big\} \,dsn(dy).  
\end{eqnarray*}

Then, taking $n = 2^k$ so that $t_j = 2^{-k}jt, \, j=0, 1,, ...,2^k$, we can 
see for each $X \in D_{x}$ that as $k \rightarrow \infty$, $S_{n1}(X)$ 
converges to the sum of the first, second and third terms in the second member
of (4.10). Thus $S_{n}(X)$ in (4.14)/(4.16) converges to the second member of (4.10),
therefore $S^{(1)}(X; x,t)$.
As a result, by the Lebesgue dominated convergence theorem the 
right-hand side of (4.15) converges to the right-hand side of (4.9). 

\medskip
II. The general case where $A(x)$ and $V(x)$ satisfy condition (3.9).

Choose a sequence $\{A_{k}\}$ 
in $C_b^{\infty}({\bf R}^d; {\bf R}^d)$ 
with $|A_{k}(x)| \leq |A(x)| $ which is 
convergent to $A(x)$ in $L^{1+\delta }_{loc}$ and 
pointwise a.e., and a sequence $\{V_{k}\}$ in 
$C_b^{\infty}({\bf R}^d; {\bf R})$ with $0\leq  V_{k}(x)\leq V(x)$ 
which is convergent to $V(x)$ in $L^{1}_{loc}$ and pointwise a.e., 
as $k\rightarrow \infty $. Then 
by (4.9), (4.10) we have 
\begin{equation}
 (e^{-t[H^{(1)}_{k}-m]}g)(x) 
= \int_{D_{x}}e^{-S_{k}(X; x,t)}g(X(t))d\lambda_x (X), 
\end{equation}
where $S_{k}(X; x,t)\,$ 
(though here with superscript ${}^{(1)}$ removed, for notational simlicity) is 
the $S^{(1)}(X; x,t)$ in (4.10) with $A_{k}$ and $V_{k}$ 
in place of $A$ 
and $V $, and $H^{(1)}_{k}$ is the selfadjoint operator associated with the form 
$h^{(1)}_{k}\equiv h^{(1)}_{A_{k}, V_{k}}$ in (2.5). We shall show both sides of 
(4.19) converge to those of (4.9) as $k\rightarrow \infty $. 

As far as the left-hand side of (4.19) is concerned, by [ITs2-93, Lemma 3.6], 
$H^{(1)}_{k}$ converges to $H^{(1)}$ in the strong resolvent sense, 
and by [Kat-76, IX, Theorem 2.16, p.504], 
$\{\exp[-t(H^{(1)}_{k}-m)]g\}^{\infty }_{k=1}$ converges to 
$\exp[-t(H^{(1)}-m)]g$, 
uniformly on each bounded $t$-interval in $[0,\infty )$, in $L^{2}$ and, if a 
subsequence is taken, pointwise a.e. 

To see convergence of the right-hand side of (4.19), we shall show 
that  $\{\exp[- S_{k} (X; x,t)]\}^{\infty }_{k=1}$ converges for 
a.e. $x$ and $\lambda _{x}$-a.e. $X$, as $k\rightarrow \infty $, 
and its limit can be written as $e^{-S^{(1)}(X; x,t)}$. Put 
$$
S_{k}(X; x,t) = \sum ^{4}_{j=1}S^{\langle j \rangle}_{k}(X; x,t), \qquad
S^{(1)}(x,t) = \sum ^{4}_{j=1}S^{\langle j \rangle}(X; x,t). 
$$
We show each $\exp[-S^{\langle j \rangle}_{k}(X; x,t)]\, (j=1,2,3,4)\,$ converges for 
$\lambda _{x}$-a.e. X. 

\medskip
(i) There exists a Borel set $K_{1}$ in ${\bf R}^{d}$ of Lebesgue measure 
zero such that $|A(x)|$ is finite and $A_{k}(x) \rightarrow  A(x)$ for 
$x \not\in K_{1}$. Then for each 
$(s,y)$ with $0<s\leq t$ and $|y| \geq 1$, 
$$
G_{1}(s,y) := \{X\in D_{x};\, X(s-)+y/2 \in  K_{1}\}
$$ 
has $\lambda _{x}$-measure zero, because 
$$
\int _{G(s,y)}d\lambda _{x}(X) = \int _{K_{1}-y/2}k_{0}(s,x-z)dz = 0.
$$ 
Therefore by the Fubini theorem 
$$
G_{1} := \{(X,s,y) \in D_{x}\times (0,t]\times \{|y| \geq 1\};\, 
X(s-)+y/2 \in  K_{1}\}
$$ 
has $[d\lambda _{x} \times dsn(dy)]$-measure zero, because 
$$
\int \int \int  I_{G_{1}}(X,s,y)d\lambda _{x}dsn(dy) = 
\int ^{t}_{0}\int _{|y| \geq 1}dsn(dy)\int _{G_{1}(s,y)}d\lambda _{x}(X) = 0, 
$$
where $I_{G_{1}}(X,s,y)$ is the indicator function 
for the set $G_{1}$. It follows again by the Fubini theorem that for 
$\lambda _{x}$-a.e. $X$, 
$$
G_{1}(X) := \{(s,y)\in (0,t]\times \{|y| \geq 1\};\, X(s-)+y/2 \in  K_{1}\}
$$ 
has $N_{X}(dsdy)$-measure zero, since 
$$
\int _{D_{x}}d\lambda _{x}(X)\int \int _{G_{1}(X)}N_{X}(dsdy) = 
\int _{D_{x}}d\lambda _{x}(X)\int \int _{G_{1}(X)}dsn(dy). 
$$
Therefore for $\lambda _{x}$-a.e. $X$, as $k\rightarrow \infty $, 
$$
A_{k}(X(s-)+y/2) \rightarrow  A(X(s-)+y/2), \qquad
N_{X}(dsdy)-\hbox{\rm a.e.},
$$
and the integral $S^{\langle 1 \rangle}_{k}(X; x,t)$ exists, being a finite sum 
because $X(s)$ has at most finitely many discontinuities $s$ with the 
jump $| X(s)-X(s-)| $ exceeding a given positive constant. By the 
Lebesgue dominated convergence theorem, for $\lambda _{x}$-a.e. $X$, 
$S^{\langle 1 \rangle}_{k}(X; x,t)$ $\rightarrow  S^{\langle 1 \rangle}(X; x,t)$ 
and hence 
$\exp[-S^{\langle 1 \rangle}_{k}(X; x,t)] 
\rightarrow  \exp[-S^{\langle 1 \rangle}(X; x,t)]$. 

\medskip
(ii) For $n>0$ let $\sigma _{n}(X) = \inf\{s>0$; $| X(s-)| >n\}$. Then for 
$\lambda_x$-a.e. $X$, $\lim_{n\rightarrow \infty }\sigma _{n}(X) = \infty $. 
For $k,\, l$ integers put $A_{kl}(x) = A_{k}(x) - A_{l}(x)$, and 
$$
G^{kl}_{2} := \{(X,s,y) \in  D_{x}\times (0,t]\times 
  \{0<| y| <1\};\, |A_{kl}(X(s-)+y/2)\cdot y| > 1\}
$$ 
and for each $X \in  D_{x}$ 
$$
G^{kl}_{2}(X) := \{(s,y) \in  (0,t]\times 
\{0<| y| <1\};\, | A_{kl}(X(s-)+y/2)\cdot y| >1\}. 
$$
The complements of $G^{kl}_{2}$ in the set $D_{x}\times (0,t]\times 
\{0<| y| <1\}$ and $G^{kl}_{2}(X)$ in the set 
$(0,t]\times \{0<| y| <1\}$ are denoted by $(G^{kl }_{2})^c$ and 
$(G^{kl}_{2}(X))^c$, respectively. 
Then we have, for $n$ fixed and for an arbitrary compact subset $K$ of 
${\bf R}^{d}$ with Lebesgue measure $|K| $, 
\begin{eqnarray*}
&&\int_{K}dx\int _{D_{x}}| S^{\langle 2 \rangle}_{k}
(X; x,t\wedge \sigma _{n}(X))-S^{\langle 2 \rangle}_{l}(X; x,t\wedge \sigma _{n}(X))| 
     d\lambda _{x}(X)\\ 
&\leq& \int_{K}dx\int _{D_{x}}
\Big| \int \int_{G_{2}^{kl }(X)}I_{[0,\sigma _{n}(X)]}(s)
A_{kl }(X(s-)+y/2)\cdot y \widetilde{N}_{X}(dsdy)\Big| d\lambda _{x}(X)\\ 
&& + \int _{K}dx\int _{D_{x}}| 
\int \int _{(G_{2}^{kl }(X))^c}I_{[0,\sigma _{n}(X)]}(s)
A_{kl}(X(s-)+y/2)\cdot y \widetilde{N}_{X}(dsdy)| d\lambda _{x}(X)\\ 
&\equiv&  \int _{K}I^{kl }_{1}dx + \int _{K}I^{kl }_{2}dx. 
\end{eqnarray*}
For $I^{kl}_{1}$ we have 
\begin{eqnarray*}
\int_{K}I^{kl }_{1}dx 
&\leq&  \int _{K}dx \int _{D_{x}}d\lambda _{x}(X)\\ 
&& \times \int \int _{G_{2}^{kl }(X)}
I_{[0,\sigma _{n}(X)]}(s)| A_{kl }(X(s-)+y/2)\cdot y|(N_{X}(dsdy)+dsn(dy))\\ 
&\leq& 2\int _{K}dx\int _{D_{x}}d\lambda _{x}(X)\int \int _{G_{2}^{kl }(X)}
I_{[0,\sigma _{n}(X)]}(s)| A_{kl }(X(s)+y/2)\cdot y| ^{1+\delta }dsn(dy)\\ 
&\leq&  2\int dx\int ^{t}_{0}\int _{0<| y| <1}dsn(dy)
\int_{|z| \leq n}| A_{kl }(z+y/2)\cdot y|^{1+\delta }k_{0}(s,x-z)dz\\ 
&\leq& 2n_{\delta }t\int _{| z| \leq n+1}| A_{kl }(z)| ^{1+\delta }dz, 
\end{eqnarray*}
with $n_{\delta} = \int_{0<|y|<1}|y|^{1+\delta}n(dy)$, where in the second 
inequality we have 
used that $|A_{kl}(X(s)+y/2)\cdot y| >$ 1 on $G^{kl }_{2}(X)$. 

For $I^{kl}_{2}$ we have 
\begin{eqnarray*}
(I^{kl }_{2})^{2} 
&=& \int_{D_{x}}d\lambda _{x}(X)\int \int_{(G_{2}^{kl }(X))^c}
I_{[0,\sigma _{n}(X)]}(s)| A_{kl }(X(s)+y/2)\cdot y| ^{2}dsn(dy)\\ 
&\leq&  \int _{D_{x}}d\lambda _{x}(X)\int \int _{(G_{2}^{kl }(X))^c}
I_{[0,\sigma _{n}(X)]}(s)| A_{kl }(X(s)+y/2)\cdot y| ^{1+\delta }dsn(dy)\\ 
&=& \int ^{t}_{0}\int _{0<| y| <1}dsn(dy)\int_{| z| \leq n}| 
A_{kl }(z+y/2)\cdot y| ^{1+\delta }k_{0}(s,x-z)dz, 
\end{eqnarray*}
where the inequality is due to that $|A_{kl }(X(s)+y/2)\cdot y|  \leq $ 1 
on $(G^{kl }_{2}(X))^c$. 
Hence 
\begin{eqnarray*}
\int_{K}I^{kl}_{2}dx 
&\leq&  |K|^{1/2}\biggl(\int (I^{kl}_{2})^{2}dx\biggr)^{1/2}\\ 
&=& |K|^{1/2}\biggl( \int dx \int ^{t}_{0}\int _{0<| y| <1}dsn(dy)\\ 
&& \qquad \qquad \times \int_{|z| \leq n}| 
A_{kl }(z+y/2)\cdot y|^{1+\delta }k_{0}(s,x-z)dz\biggr)^{1/2}\\ 
&\leq&  (| K| n_{\delta }t)^{1/2}
\biggl( \int _{| z| \leq n+1}| A_{kl }(z)| ^{1+\delta }dz\biggr) ^{1/2}. 
\end{eqnarray*}
Thus we have 
\begin{eqnarray*}
&& \int_{K}dx\int_{D_{x}}| 
  S^{\langle 2 \rangle}_{k}(X; x,t\wedge \sigma _{n}(X))
  -S^{\langle 2 \rangle}_{l }(X; x,t\wedge \sigma _{n}(X))| 
         d\lambda _{x}(X)\\ 
&\leq&  \int_{K}I^{kl}_{1}dx + \int _{K}I^{kl }_{2}dx\\ 
&\leq& 2n_{\delta }t\int _{| z| \leq n+1}|A_{kl }(z)|^{1+\delta }dz 
+ (| K| n_{\delta }t)^{1/2}\biggl( \int _{|z| \leq n+1}
| A_{kl }(z)| ^{1+\delta }dz\biggr)^{1/2}, 
\end{eqnarray*}
which tends to zero as $k, l  \rightarrow  \infty $. Since $K$ is 
arbitrary, it can be seen, by passing to a subsequence, for a.e. $x$, that 
as $k\rightarrow \infty $, 
$\{S^{\langle 2 \rangle}_{k}(X; x,t\wedge \sigma _{n}(X))\}^{\infty }_{k=1}$ 
converges to a 
limit in $L^{1}$ with respect to $\lambda _{x}$ 
and, by passing to a subsequence, for $\lambda _{x}$-a.e. X. This limit is 
what is to be denoted by $S^{\langle 2 \rangle}(X; x,t\wedge \sigma _{n}(X))$. 
Further, since 
$\lim_{n\rightarrow \infty }\sigma _{n}(X) = \infty $ 
for $\lambda _{x}$-a.e. $X$, we see that 
$\{S^{\langle 2 \rangle}_{k}(X; x,t)\}^{\infty }_{k=1}$ converges to a limit for 
$\lambda _{x}$-a.e. $X$, which is to be denoted by 
$S^{\langle 2 \rangle}(X; x,t)$, and hence 
$\exp[-S^{\langle 2 \rangle}_{k}(X; x,t)] 
\rightarrow  \exp[-S^{\langle 2 \rangle}(X; x,t)]$. 

\medskip
(iii) We can show with the theory of singular integrals that 
$$
a(x) := {\rm p.v.}\int _{0<| y| <1}A(x+y/2)\cdot y \, n(dy)
$$ 
exists pointwise a.e. in $x$, while 
$$
a_{k}(x) := \int _{0<| y| <1}(A_{k}(x+y/2)-A_{k}(x))\cdot y \, n(dy) 
= {\rm p.v.}\int _{0<| y| <1}A_{k}(x+y/2)\cdot y \, n(dy)
$$ 
exists for every $x$, and that as $k\rightarrow \infty $, $a_{k}(x)$ 
converges to $a(x)$ in $L^{1+\delta }_{loc}$. 
With the same $\sigma _{n}(X)$ as in (ii), we have, for $n$ fixed, 
\begin{eqnarray*}
&&\int_{{\bf R}^d} dx\int _{D_{x}}| S^{\langle 3 \rangle}_{k}
(X; x,t\wedge \sigma _{n}(X)) - S^{\langle 3 \rangle}_{l}
(X; x,t\wedge \sigma _{n}(X))| ^{1+\delta } d\lambda _{x}(X)\\ 
&=& \int dx\int d\lambda _{x}(X)\Big| 
\int ^{t}_{0}I_{[0,\sigma _{n}(X)]}(s)[a_{k}(X(s))-a_{l }(X(s))]ds
   \Big|^{1+\delta }\\ 
&\leq& \int dx\int d\lambda _{x}(X) 
 \biggl(\int ^{t}_{0}ds\biggr)^{\delta} 
\int ^{t}_{0}I_{[0,\sigma _{n}(X)]}(s)
     | a_{k}(X(s))-a_{l }(X(s))|^{1+\delta }ds\\ 
&\leq&  
t^{\delta }\int dx\int ^{t}_{0}ds\int _{| z| \leq n}
  | (a_{k}(z)-a_{l }(z))| ^{1+\delta }k_{0}(s,x-z)dz\\ 
&\leq& t^{1+\delta }\int _{| x| \leq n}
| (a_{k}(x)-a_{l }(x)| ^{1+\delta }dx 
\rightarrow  0, \qquad k,l  \rightarrow  \infty, 
\end{eqnarray*}
where in the first inequality we have used the H\"{o}lder inequality. 
Similarly to (ii), it can be seen, by passing to a subsequence, for 
a.e. $x$, that as $k \rightarrow  \infty $, 
$\{S^{\langle 3 \rangle}_{k}(X; x,t\wedge 
   \sigma _{n}(X))\}^{\infty }_{k=1}$ converges to 
 a limit in $L^{1+\delta }$ with respect to $\lambda _{x}$ and, 
 by passing to a subsequence, for $\lambda _{x}$-a.e. X. 
This limit is what is to be denoted by 
$S^{\langle 3 \rangle}(X; x,t\wedge \sigma _{n}(X))$. 
Further, since $\lim_{n\rightarrow \infty }\sigma _{n}(X) = \infty $ for 
$\lambda _{x}$-a.e. $X$, we see that 
$\{S^{\langle 3 \rangle}_{k}(X; x,t)\}^{\infty }_{k=1}$ converges to a limit for 
$\lambda_{x}$-a.e. $X$, which is to be 
denoted by $S^{\langle 3 \rangle}(X; x,t)$, and hence 
$\exp[-S^{\langle 3 \rangle}_{k}(X; x,t)] 
\rightarrow  \exp[-S^{\langle 3 \rangle}(X; x,t)]$. 

\medskip
(iv) The proof will proceed in the same way as the proof of [S2-79/05,  
II, Theorem 6.2, p.51]. We may suppose that $V_{k}(x) \uparrow  V(x)$ 
poitwise a.e. There exists a Borel set $K_{4}$ in ${\bf R}^{d}$ of Lebesgue 
measure zero such 
that $V_{n}(x) \uparrow  V(x)$ for $x \not\in K_{4}$. Then for 
$0<s\leq t$, 
$$
G_{4}(s,y) = \{X\in D_{x};\, X(s) \in K_{4}\}
$$ 
has $\lambda_{x}$-measure zero. Therefore by the Fubini theorem 
$$
G_{4} = \{(X,s) \in  D_{x}\times (0,t];\, X(s) \in  K_{4}\}
$$ 
has $[d\lambda _{x} \times ds]$-measure zero, so that for 
$\lambda _{x}$-a.e. $X$, 
$$
G_{4}(X) = \{s\in (0,t];\, X(s) \in  K_{4}\}
$$ 
has Lebesgue measure zero. It follows by the monotone 
covergence theorem that for $\lambda _{x}$-a.e. $X$, 
$S^{\langle 4 \rangle}_{n}(X; x,t) 
  \rightarrow  S^{\langle 4 \rangle}(X; x,t)$ and 
hence $\exp[-S^{\langle 4 \rangle}_{n}(X; x,t)] 
\rightarrow  \exp[-S^{\langle 4 \rangle}(X; x,t)]$. 
This proves Theorem 4.1.  \qed

\bigskip\noindent
(2) Next we come to the case for $H^{(2)} := H^{(2)}_A + V$ 
in Definition 3.3 with condition (3.9) on $A(x)$ and $V(x)$. 

\begin{thm} {\rm [IfMP1-07, 2-08, 3-10]} 
The same hypothesis as in Theorem 4.1.
\begin{eqnarray}
(e^{-t[H^{(2)}-m]}g)(x) 
 &\!\!=\!\!& \int_{D_x([0,\infty)\rightarrow {\bf R}^d)} e^{-S^{(2)}(X; x,t)} 
        g(X(t))\,d\lambda_x (X), \\ 
S^{(2)}(X; x,t)
&\!\!=\!&\!\!\!\, i\int_0^{t+}\int_{|y|>0} 
       \Big(\int_0^1 A(X(s-)\!+\!\theta y)d\theta\Big)\!\cdot\! y\, 
               \widetilde{N}_X(dsdy) \\
{}\!\!\!&+&\!\!\!\, i\int_0^{t}\int_{|y|>0}
      \!\Big[\!\int_0^1\! A(X(s)\!+\!\theta y)d\theta -A(X(s))\Big]
    \!\cdot\! y dsn(dy) \!+\!\!\int_0^t V(X(s))ds. \nonumber
\end{eqnarray}
\end{thm}

\medskip
The proof of Theorem 4.3 can be done in exactly the same way as that 
of Theorem 4.1. Indeed, we have only to replace 
$A(X(s-)+\frac{y}{2})\!\cdot\! y$ by 
$(\int_0^1 A(X(s-)+\theta y)d\theta) \!\cdot\! y\,$. 
We will not repeat it here. 

\bigskip\noindent
(3) Finally, we consider the case for the operator defined, 
in Definition 3.4,
 with the square root of a nonnegative selfadjoint operator, 
$H^{(3)} := H^{(3)}_A + V$.

On the one hand, we can determine by functional analysis, namely, 
by theory of fractional powers (e.g. Yosida [Y, Chap. IX, 11, pp.259--261]) 
$e^{-t[H^{(3)}_A-m]}$  from the nonnegative selfadjoint operator
$S:= (-i\nabla-A(x))^2+m^2 =: 2H_A^{NR}+m^2$, where
$H_A^{NR} = \frac1{2}(-i\nabla-A(x))^2$ is the magnetic nonrelativistic 
Schr\"odinger operator with mass $1$ and without scalar potential. 
Indeed, we have
\begin{eqnarray}
e^{-t[H_A^{(3)}-m]}g &=&\left\{
\begin{array}{rl}
e^{mt} \int_0^{\infty} f_t(\kappa)e^{-\kappa S} g\, d\kappa, &\quad t>0,\\
 0, &\quad t=0
\end{array}\right. \nonumber\\
f_t(\kappa) &=& \left\{
\begin{array}{rl}
  (2\pi i)^{-1} \int_{\sigma-i\infty}^{\sigma+i\infty}
    e^{z\kappa -tz^{1/2}}dz, &\quad \kappa \geq 0,\\
0, &\quad \kappa <0 \quad(\sigma >0).
\end{array}\right.
\end{eqnarray}
This equation (4.22) may provide a kind of path integral representation for 
$e^{-t[H_A^{(3)}-m]}g$ with the Wiener measure $\mu_x^{1}$ 
corresponding to mass $1$ [cf. $\mu_x^{m} \equiv \mu_x$ in (4.29)]: 
\begin{eqnarray*}
&\!\!&\!\!(e^{-t[H_A^{(3)}-m]}g)(x) \\
\!\!&\!=\!&\!\! e^{mt}\int_0^{\infty} d\kappa{}\,{} f_t(\kappa)
              e^{-\kappa m^2}\\
&\!\!&\!\!\times 
  \int_{C_x([0,\infty)\rightarrow {\bf R}^d)} \,
 \!e^{-[i\int_0^{2\kappa} A(B(s))\circ dB(s)+\int_0^{2\kappa}V(B(s))ds]}
   g(B(2\kappa m))\, d\mu_x^{1}(B),
\end{eqnarray*}
though with an undesirable extra $d\kappa$-integral,
by substituting the Feynman--Kac--It\^o formula (4.2) 
with $V=0$, i.e. for $e^{-t[H_A^{(3)}-m]}g$ with $t=2\kappa$ 
into $e^{-\kappa (S-m^2)}= e^{-2\kappa H_A^{NR}}$ in the integrand of 
(4.22). 

Then if we would use this further to represent $e^{-t[H^{(3)}-m]}g$ 
for $V \not= 0$, we might apply the Trotter--Kato product formula 
\begin{equation}
e^{-t[H^{(3)}-m]} = 
\hbox{s-}\!\!\lim_{n\rightarrow \infty}
\big(e^{-(t/n)[H_A^{(3)}-m]}e^{-(t/n)V}\big)^n,
\end{equation}
for the sum $H^{(3)}-m = (H_A^{(3)}-m)+V$ to express the semigroup 
$e^{-t[H^{(3)}-m]}$ as a ``limit",
where convergence of the right-hand side usually takes place  
in strong operator topology as indicated. 
However it is not clear whether this procedure could further yield 
a path integral representation for $e^{-t[H^{(3)}-m]}g$.

In passing, let us insert a comment on the convergence 
of Trotter--Kato product formula (4.23). It is now known 
that the convergence takes place even {\it in operator norm}, 
so long as the operator sum $(H_A^{(3)}-m)+V$ is selfadjoint on
the common domain $D[H_A^{(3)}]\cap D[V]$ 
by the recent results in Ichinose--Tamura [IT1-01], 
Ichinose--Tamura--Tamura--Zagrebnov [ITTaZ-01] and also
even pointwise convergence of the integral kernels in [IT2-04, 3-06] 
(cf. [I8-99]). 

On the other hand, it does not seem possible to represent $e^{-t[H^{(3)]}-m]}g$
by path integral through directly applying L\'evy process as we saw 
in the cases for $e^{-t[H^{(1)}-m])}g$ and $e^{-t[^{(2)}-m]}g$,
because 
$H_A^{(3)}$ does not seem to be explicitly expressed by 
a pseudo-differential operator corresponding to a certain tractable symbol.
It was in this situation that the problem of path integral 
representation for 
$e^{-t[H^{(3)]}-m]}g$ was studied first by DeAngelis--Serva [DeSe-90] 
and DeAngelis--Rinaldi--Serva [DeRSe-91]
with use of {\it subordination /time-change} of Brownnian motion,
and then by Nagasawa [N1-96, 2-97, 3-00].
Recently it has been more extensively studied 
by Hiroshima--Ichinose--L\H{o}rinczi [HILo1-12, 2-12] (cf. [LoHB-11]) 
not only for the magnetic relativistic Schr\"odinger operator $H_A^{(3)}$
but also for Bernstein functions of the magnetic nonrelativistic
Schr\"odinger operator and  even with spin. 
In this connection, the problem on nonrelativistic limit was studied in [I1-87],[Sa1-90], [N1-97].

To proceed, let us explain about subordination 
(e.g. [Sa2, Chap.6, p.197], [Ap-04/09, 1.3.2, p.52]).
Subordination is a transformation, through random time change, 
of a stochastic process to a new one which is a non-decreasing L\'evy process 
independent of the original one, what is called {\it subordinator}.
The new process is said to be {\it subordinate} to the original one. 

As the original process, take $B^1(t)$, the one-dimensional 
standard Brownian motion, 
so that $B^1 \equiv B^1(\cdot)$ is a function belonging to the space
$C_0([0,\infty)\!\!\rightarrow\!\!{\bf R})$ of real-valued continuous functions
on $[0,\infty)$ satisfying $B^1(0) =0$ and
$$  
e^{-t\frac{\xi^2}{2}} 
= \int_{C_0([0,\infty)\rightarrow{\bf R})} e^{i\xi B^1(t)} d\mu_0^S(B^1),
$$
where $\mu_0^S$ is the Wiener measure on $C_0([0,\infty)\!\!\rightarrow\!\!{\bf R})$. 
Let $m \geq 0$, and for each $B^1$ and $t\geq 0$, put
\begin{equation}
 T(t) \equiv T(t,B^1) 
 := \inf \{\, s > 0\,;\, B^1(s)+m\,s = t\}.
\end{equation}
Then $T \equiv T(\cdot)$ is a monotone, non-decreasing function on $[0,\infty)$
with $T(0)=0$, belonging to $D_0([0,\infty)\!\!\rightarrow\!\! {\bf R})$
and so becoming a one-dimensional L\'evy process, called 
{\it inverse Gaussian subordinator} for $m>0$ and {\it L\'evy subordinator} for $m=0$. 
This correspondence defines a map $\hat{T}$ of
$C_0([0,\infty)\!\!\rightarrow\!\!{\bf R})$ into 
$D_0([0,\infty)\!\!\rightarrow\!\!{\bf R})$
by $\hat{T} B^1(\cdot) = T(\cdot,B^1)$. Let $\nu_0$ be the probability measure
 on $D_0([0,\infty)\!\!\rightarrow\!\! {\bf R})$ defined by 
$\nu_0(G) = \mu_0^S(\hat{T}^{-1}G)$ first for cylinder subsets  
$G \subset D_0([0,\infty)\!\!\rightarrow\!\! {\bf R})$ and then extended 
to more general subsets.

\begin{prp} {\rm (e.g. [Ap-04/09, Example 1.3.21, p.54, 
and Exercise 2.2.10, p.96; cf. Theorem 2.2.9, p.95])} 
The probability measure $\nu_0$ satisfies
 \begin{equation} 
 e^{-t[\sqrt{2\sigma+m^2}-m]} 
= \int_{D_0([0,\infty)\rightarrow{\bf R})}e^{-T(t)\sigma} d\nu_0(T),
\qquad \sigma \geq 0.
\end{equation}
\end{prp}

\bigskip\noindent
{\it Proof}. The proof will be not selfcontained, and need some basic facts 
about {\it martingale} and {\it stopping time} 
(e.g. [IkW2-81/89], [Sa2-99], [Ap-04/09], [DvC-00], [LoHB-11]). 
$T(t,B^1)=T(t)$ is a stopping time and then
$B^1(T(t))$ is a {\it stopped random variable} belonging to 
$D_0([0,\infty)\!\!\rightarrow\!\! {\bf R})$.
Then we see that, for $\theta \in {\bf R}$,
$$
M_{\theta}(t) := e^{\theta B^1(t)-\frac12\theta^2t}
$$ is a continuous 
martingale with respect to the natural filtration of Brownian motion $B^1(t)$. 
Further, 
$$M_{\theta}((\hat{T} B^1)(t)\wedge n) 
= e^{B^1(T(t)\wedge n)- \frac12\theta^2 T(t)\wedge n}
$$
is also a martingale. Then by {\sl Doob's optional stopping theorem} 
[Ap-04/09, Theorem 2.2.1, p.92], for each $t>0$, positive integer $n$
and $\theta \geq 0$, we have
\begin{eqnarray*}
\int_{C_0([0,\infty)\rightarrow {\bf R})}
M_{\theta}((\hat{T}B^1)(t)\wedge n) d\mu_0^S(B^1)
&=& \int_{C_0([0,\infty)\rightarrow {\bf R})}
  M_{\theta}((\hat{T}B^1)(0)\wedge n) d\mu_0^S(B^1)\\
&=& \int_{C_0([0,\infty)\rightarrow {\bf R})}e^{\theta B^1(0)}d\mu_0^S(B^1)=1.
\end{eqnarray*}
For each positive integer $n$ and $t>0$, put
$\Omega_{n,t} := \{B^1 \in 
C_0([0,\infty)\rightarrow {\bf R})\,;\, T(t) =(\hat{T}B^1)(t)\leq n\}$.
Then
\begin{eqnarray*}
&&\int_{C_0([0,\infty)\rightarrow {\bf R})}
 M_{\theta}((\hat{T}B^1)(t)\wedge n)  d\mu_0^S(B^1)\\
&=& \int_{\Omega_{n,t}}M_{\theta}((\hat{T}B^1)(t)\wedge n) d\mu_0^S(B^1)
  + \int_{{\Omega_{n,t}}^c}M_{\theta}((\hat{T}B^1)(t)\wedge n) d\mu_0^S(B^1).
\end{eqnarray*}
Since, for $B^1(t)$, $(\hat{T} B^1)(t) =T(t) >n$ implies 
$B^1(n) < t - m n$, so that
$$
\int_{{(\Omega_{n,t})}^c}M_{\theta}((\hat{T}B^1)(t)\wedge n) d\mu_0^S(B^1)
\leq e^{-\frac12\theta^2 n}\int_{{\Omega_{n,t}}^c}
e^{\theta B^1(n)}d\mu_0^S(B^1)
\leq e^{-\frac12\theta^2n +\theta (t- mn)},
$$
which, for $\theta >0$, tends to zero as $n\rightarrow \infty$. It follows 
by the monotone convergence theorem and (4.24) that
\begin{eqnarray*}
1 &=& 
\int_{C_0([0,\infty)\rightarrow {\bf R})}
    M_{\theta}((\hat{T} B^1)(t)\wedge n) d\mu_0^S(B^1)\\
&=& \lim_{n\rightarrow \infty}
\int_{\Omega_{n,t}}M_{\theta}((\hat{T}B^1)(t)\wedge n) d\mu_0^S(B^1)
= \int_{C_0([0,\infty)\rightarrow {\bf R})}
    M_{\theta}((\hat{T}B^1)(t)) d\mu_0^S(B^1)\\
&=& \int_{C_0([0,\infty)\rightarrow {\bf R})}
e^{\theta [t-m(\hat{T} B^1)(t)]-\frac12\theta^2(\hat{T} B^1)(t)}
     d\mu_0^S(B^1),
\end{eqnarray*}
whence
\begin{eqnarray*}
e^{-\theta t} 
&=& \int_{C_0([0,\infty)\rightarrow {\bf R})}
e^{-\frac12\theta(\theta +2m)(\hat{T} B^1)(t)} d\mu_0^S(B^1)\\
&=& \int_{D_0([0,\infty)\rightarrow {\bf R})}
e^{-\frac12\theta(\theta +2m)T(t)} d\nu_0(T).
\end{eqnarray*}
Taking $\theta = \sqrt{2\sigma+m^2}-m$ yields the result, 
showing Proposition 4.4. \qed

\bigskip
This proposition implies that the characteristic function of the measure 
$\nu_0$ is given by 
\begin{eqnarray}
 e^{-tV(\rho)} 
\!\!&\!=\!&\!\! 
  \int_{D_0([0,\infty)\rightarrow{\bf R})}e^{iT(t)\rho} d\nu_0(T),
\qquad \rho \in {\bf R}\,, \nonumber\\
V(\rho) 
\!\!&\!=\!&\!\! 
  \frac{\sqrt{m^4+4\rho^2}-m^2}{\sqrt{2}[(m^2+\sqrt{m^4+4\rho^2})^{1/2}+\sqrt{2}m]}
- \frac{\sqrt{2}\rho}{(m^2+\sqrt{m^4+4\rho^2})^{1/2}}\,i \nonumber\\
\!\!&\!=\!&\!\! 
  \frac{2\sqrt{2}\rho^2}{[(m^2+\sqrt{m^4+4\rho^2})^{1/2}+\sqrt{2}m]
   (m^2+\sqrt{m^4+4\rho^2})} \nonumber\\
 &&\qquad\qquad\qquad\qquad 
  -\frac{\sqrt{2}\rho}{(m^2+\sqrt{m^4+4\rho^2})^{1/2}}\,i \,.
\qquad\quad
\end{eqnarray}
To see this, first analytically extend 
$\sqrt{2\sigma+m^2}$ to the right-half complex plane 
$z := \sigma +i\rho,\,\, \sigma>0,\, \rho \in {\bf R}$, and next we have 
$V(-\rho) = \lim_{\sigma\rightarrow +0} \sqrt{2(\sigma + i\rho)+m^2}-m$,
of which the right-hand side is to be calculated. Then (4.26) follows 
with $\rho$ replaced by $-\rho$.
[cf. Using a subordinator $T(t)$ slightly different from (4.24), in 
[I9-12, (4.18), (4.19), (4.20), p.335] there are given a little different 
formulas corresponding to (4.25) and (4.26), $V(\rho)$. However, 
it contains an error; ``$\,\rho^2\,$" in the expression for $V(\rho)$ 
there should be replaced by ``$\,4\rho^2\,$".]

\medskip
Now we are in a position to give a path integral representation for 
$e^{-t[H^{(3)}-m]}g$.  

\begin{thm} 
\begin{eqnarray}
(e^{-t[H^{(3)}-m]}g)(x)
\!\!&\!=\!&\!\!
\int\int_{\stackrel{\scriptstyle C_x([0,\infty)\rightarrow {\bf R}^d)}
{\times D_0([0,\infty)\rightarrow{\bf R})}} \,e^{-S^{(3)}(B,T; x,t)}
     g(B(T(t))) \,d\mu_x^{1}(B) d\nu_0(T),\\ 
S^{(3)}(B,T; x,t)
\!\!&\!=\!&\!\! i\int_0^{T(t)} \!\!A(B(s))\,dB(s)
       +\frac{i}{2}\int_0^{T(t)}\!\! \hbox{\rm div} A(B(s))ds 
        + \int_0^t \!\!V(B(T(s)))ds,\nonumber\\
\!\!&\!\equiv\!&\!\! 
    i\int_0^{T(t)} A(B(s))\circ dB(s)+ \int_0^tV(B(T(s)))ds,
\end{eqnarray}
\noindent
where $\mu_x^{1}$ stands for the Wiener measure 
$\mu_x^{m} \equiv \mu_x$  on 
$C_x([0,\infty)\rightarrow{\bf R}^d)$ with characteristic function
\begin{equation}
 \exp\Big[-t\frac{|\xi|^2}{2}\Big]  
=\int_{C_x([0,\infty)\rightarrow{\bf R}^d)}e^{i(B(t)-x)\cdot \xi} 
     d\mu_x^{1}(B),
\end{equation}
with mass taken as $m=1$. 
\end{thm}

\medskip\noindent
{\sl Remark}. We note that for every pair 
$(B,T) \in C_x([0,\infty)\rightarrow{\bf R}^d)
          \times D_0([0,\infty)\rightarrow{\bf R})$ 
the path $B(T(s))$ in this theorem belongs to
$D_x([0,\infty)\rightarrow{\bf R}^d)$, because it is right-continuous in 
$s \in [0,\infty)$ and has left-hand limit. The characteristic function 
of the product $\mu_x^{1} \times \nu_0$ of the probability measures is 
calculated with (4.29) as
\begin{eqnarray}
&\!\!&\!\! 
\int\int_{\stackrel{\scriptstyle C_x([0,\infty)\rightarrow {\bf R}^d)}
{\times D_0([0,\infty)\rightarrow{\bf R})}} \,e^{i(B(T(t))-x)\cdot \xi} 
\,d\mu_x^{1}(B) d\nu_0(T) \nonumber\\
&\!=\!&\!\! \int_{D_0([0,\infty)\rightarrow{\bf R})} \,
\exp\Big[-T(t)\frac{|\xi|^2}{2}\Big]\, d\nu_0(T) 
= e^{-t[\sqrt{|\xi|^2+m^2}-m]},
\end{eqnarray}
thus coinciding with (4.4), the characteristic function of the measure
$\lambda_x$. 
This implies that these two processes on the two different probability spaces
$\big(C_x([0,\infty)\rightarrow{\bf R}^d)
   \times D_0([0,\infty)\rightarrow{\bf R}),\, 
      \mu_x^{1} \times \nu_0\big)$
and
$\big(D_x([0,\infty)\rightarrow{\bf R}^d),\, \lambda_x\big)$
are identical in law, i.e. have the same finite-dimensional 
distributions, in fact, for $0< t_1< t_2< \dots < t_n <\infty$ 
and $n=1, 2, 3, \dots$,
\begin{eqnarray*}
&\!\!&\!\! 
\int\int_{\stackrel{\scriptstyle C_x([0,\infty)\rightarrow {\bf R}^d)}
{\times D_0([0,\infty)\rightarrow{\bf R})}} \,
e^{i[(B(T(t_1))-x)\cdot \xi_1 +(B(T(t_2))-x)\cdot \xi_2 
    +\cdots + (B(T(t_n))-x)\cdot \xi_n]} 
\,d\mu_x^{1}(B) d\nu_0(T) \\
&=& \int_{D_x([0,\infty)\rightarrow{\bf R}^d)}
e^{i[(X(t_1)-x)\cdot \xi_1 +(X(t_2)-x)\cdot \xi_2 
    +\cdots + (X(t_n)-x)\cdot \xi_n]}\, d\lambda_x(X) \\
&=& e^{-t_1[\sqrt{|\xi_1+\xi_2+\cdots+\xi_n|^2+m^2}-m]}
    e^{-(t_2-t_1)[\sqrt{|\xi_2+\cdots+\xi_n|^2+m^2}-m]}\cdot \cdots \cdot \\
 &&\qquad\qquad\qquad\qquad\qquad\qquad\qquad\qquad\qquad\qquad\qquad \times
    e^{-(t_n-t_{n-1})[\sqrt{|\xi_n|^2+m^2}-m]}.
\end{eqnarray*}%
Therefore the former process may also be considered basically a L\'evy process,
but it is not clear whether one can rewrite the right-hand side of (4.27)
as a process on the probability space 
$\big(D_x([0,\infty)\rightarrow{\bf R}^d),\, \lambda_x\big)$,
replacing the function $S^{(3)}(B,T; x,t)$ of $\, B,\, T,\,x,\,t$ in (4.28)
with some function else $S^{(3)}(X; x,t)$ of $X,\,x,\,t$ appropriately written 
in terms of the L\'evy space path $X$ in $D_x([0,\infty)\rightarrow{\bf R}^d)$. 

\bigskip\noindent
{\it Proof of Theorem 4.5}. We give only a sketch. 
The detail is referred to [IHLo1-12]. 
We use Proposition 4.4 and the Feynman--Kac--It\^o formula
(4.2). Note that $H_A^{(3)} = \sqrt{2H_A^{NR}+m^2}$ where
$H_A^{NR} = \frac12(-i\nabla-A(x))^2$ is the magnetic nonrelativistic Schr\"odinger
operator with mass $1$.
By the spectral theorem for the nonnegative selfadjoint operator 
$H_A^{NR}$, we have 
$H_A^{NR} = \displaystyle{\int_{\hbox{\rm\scriptsize{Spec}}(H_A^{NR})}} 
\sigma \,dE(\sigma)$, where $E(\cdot)$ is the spectral measure associated with
$H_A^{NR}$. Then for $f,\, g \in L^2({\bf R}^d)$
$$
(f, e^{-t[H_A^{(3)}-m]}g) = 
\int_{\hbox{\rm \scriptsize{Spec}}(H_A^{NR})} 
   e^{-t[\sqrt{2\sigma+m^2}-m]} \,(f, dE(\sigma)g).
$$
Here we are using the {\sl physicist's inner product} $(f, g)$, 
which is anti-linear in $f$ and linear in $g$. 
By Propositopn 4.4 and again by the spectral theorem, 
\begin{eqnarray*}
(f, e^{-t [H_A^{(3)}-m]}g)  
&=& \int_{\hbox{\rm \scriptsize{Spec}}(H_A^{NR})}  
   \int_{D_0([0,\infty)\rightarrow {\bf R})}e^{-T(t)\sigma} d\nu_0(T) \,
   (f, dE(\sigma)g)\\
&=& \int_{D_0([0,\infty)\rightarrow {\bf R})} 
      (f, e^{-T(t)H_A^{NR}} g)\, d\nu_0(T).
\end{eqnarray*}
Applying the Feynman--Kac--It\^o formula (4.2) 
(for the case $V=0$) to $e^{-T(t) H_A^{NR}}g$ in the third member, we have
\begin{eqnarray*}
\!\!&&\! (f,e^{-t  [H_A^{(3)}-m]}g) \\
\!&\!\!\!=\!\!\!&\!
  \int_{D_0([0,\infty)\rightarrow {\bf R})}\!\!\!\!\! d\nu_0(T)
 \!\!\int_{{\bf R}^d} \!\!\!dx \overline{f(\!B(0))}\!\!\!
 \int_{C_x([0,\infty)\rightarrow {\bf R}^d)}
  \!\!\!\!e^{-i\int_0^{T(t)}\!\!A(\!B(s))\circ dB(s)}\!g(\!B(T(t))) 
d\mu_x^{1}(\!B)\\
\!&\!\!=\!\!&\!\int_{{\bf R}^d} dx \overline{f(x)}
 \!\!\int\int_{\stackrel{\scriptstyle C_x([0,\infty)\rightarrow {\bf R}^d)}
{\times D_0([0,\infty)\rightarrow{\bf R})}}\,
   e^{-i\int_0^{T(t)} \!A(B(s))\circ dB(s)} g(\!B(\!T(t))) 
\,d\mu_x^{1}(B)d\nu_0(T),
\end{eqnarray*}
where note $B(0)=x$.
This proves the assertion when $V= 0$.

When $V\not= 0$, with  partition of $[0,t]$:   
$0=t_0<t_1< \cdots < t_n=t,\,\, t_j-t_{j-1}=t/n,$
we can express $e^{-t  [H^{(3)}-m]}g = e^{-t[(H_A^{(3)}-m)+V]}$ 
by the Trotter--Kato formula (4.23).
Rewrite the product of these $n$ operators
by path integral with respect to the product of two probability measures 
$\nu_0(T) \cdot\mu_x^{1}(B)$ and note that 
$T(0)\!=\!T(t_0)\!=\!0,\,  B(0)\!=\!B(T(t_0))\!=\!x$, then  we have 
\begin{eqnarray*}
&&\!\!\!
(f, \big(e^{-(t/n)[H_A^{(3)}-m]}e^{-(t/n)V}\big)^ng)\\
&\!\!=\!\!& \int_{{\bf R}^d} dx
\int_{D_0([0,\infty)\rightarrow {\bf R})}\!\! d\nu_0(T)
\int_{C_x([0,\infty)\rightarrow {\bf R}^d)} \overline{f(B(0))}\\
&&\qquad\qquad\qquad
\times e^{-i\sum_{j=1}^n \int_{T(t_{j-1})}^{T(t_j)} A(B(s))\circ dB(s)}
e^{-\sum_{j=1}^n  V(B(T(t_{j}))\frac{t}{n}} g(B(t_n))\,d\mu_x^{1}(B).
\end{eqnarray*}
We see that, as $n\rightarrow \infty$, the left-hand side converges to 
$(f, e^{-t  [H^{(3)}-m]}g)$, and the Lebesgue theorem shows 
the right-hand side converges as integral by the product measure
$dx \times \nu_0(T)\times \mu_x^{1}(B)$, so that we obtain
\begin{eqnarray*}
&&\!\!\!
(f, \big(e^{-t[H^{(3)}-m]}g)\\
&\!\!=\!\!& 
\int_{{\bf R}^d} dx \overline{f(x)}
 \!\!\int\int_{\stackrel{\scriptstyle C_x([0,\infty)\rightarrow {\bf R}^d)}
{\times D_0([0,\infty)\rightarrow{\bf R})}}\,
   e^{-S^{(3)}(B,T; x,t)}g(\!B(\!T(t)))\,d\mu_x^{1}(B)d\nu_0(T).
\end{eqnarray*}
Hence or similarly we can also get (4.27)/(4.28) 
with $f$ removed in the above inner products. 
\qed


\subsection{Heuristic derivation of path integral formulas}

After a brief introduction to {\sl path integral}, we discuss, 
for the solution of the imaginary-time magnetic relativistic Schr\"odinger 
equation (4.3), how to heuristically derive its  
path integral formulas (4.9)/(4.10) in Theorem 4.1, (4.20)/(4.21) 
in Theorem 4.3 and (4.27)/(4.28) in Theorem 4.5.

\medskip\noindent 
4.2.1. What is {\sl path integral} ? 
 
It is a fabulous technique invented by Feynman in his Princeton 
1942 thesis (see [Fey2-05]) and his 1948 paper [Fey1-48] 
to give alternative formulation of quantum mechanics. 
Its like has never been made before or since.
In fact, though it is not mathematically rigorous, 
because of the universality of its idea, 
it has now come to prevail over all the domains in quantum physics.  
It is interesting to note, as he himself wrote in [48],
that he came to the idea, ``suggested by some of Dirac's remarks 
([Di1-33, 2-35], [Di3-45]) concerning the relation of classical action to 
quantum mechanics." 
It is a special kind of {\sl functional integral} like 
\begin{equation}
 \int e^{\frac{i}{\hbar}S(X)} {\cal D}[X]
\end{equation}
on space of paths $X: [0,t]\ni s \mapsto X(s) \in {\bf R}^d$
with respect to a `measure' ${\cal D}[X]$ 
on the space of these paths, 
where we are restoring the physical constant 
$\hbar = \frac{h}{2\pi} \,\  (h>0$: {Planck's constant}).
$S(X)$ is time integral of the {\sl Lagrangian} $L(X(s), \dot{X}(s))$
where $\dot{X}(s) = \frac{d}{ds}X(s)$: 
$$
S(X) = \int_0^t L(X(s), \dot{X}(s)) ds,
$$
which is an important quantity in classical mechanics, 
called {\sl action} along the path $X$, having 
physical dimension of Planck's constant 
$h$ so that $\frac{S(X)}{\hbar}$ becomes dimensionless.

We have in mind the nonrelativistic-quantum-mechanical motion of a  particle 
in space ${\bf R}^d$ under influence of the scalar potential $V(x)$. 
In the previous sections, we let the particle have the special mass $m=1$, but 
in this section, for a while, assume it to have general mass $m>0$ 
so that we can see where $m$ appears in the following description of its dynamics.  
Thus consider the Cauchy problem for the nonrelativistic Schr\"odinger equation 
for this particle with initial data $\psi(x,0)=f(x)$:
\begin{equation}
   i \hbar \frac{\partial}{\partial t} \psi (x,t)
   = \Big[- \frac{\hbar^2}{2m} \Delta + V(x)\Big] \psi (x,t), 
   \quad t > 0, \quad x  \in {\bf R}^d. 
\end{equation}
The solution is expressed as 
$$
\psi (x,t) = \int K(x,t; y,0) f(y) dy,
$$ 
where $K(x,t; y,0)$ is integral kernel, called
{\it fundamental solution} or {\it propagator}.

Feynman writes down this important quantity $K(x,t; s,y)$ as
an `integral' 
\begin{equation}
K(x,t; y,0) = \int_{\{X: \, X(0) = y, X(t) = x \}} e^{\frac{i}{\hbar} S(X)}
  {\cal D}[X],
\end{equation} 
where in the present case 
$L(X(s), \dot{X}(s)) = \frac{m}{2} \dot{X}(s)^2 - V(X(s))$, so that 
$S(X)$ is given by 
\begin{equation}    
 S(X) = \int_0^t \Big[\frac{m}{2} \dot{X}(s)^2 - V(X(s))\Big] ds.
\end{equation}
${\cal D}[X]$ stands for a uniform `measure', if it exists,  
on  the space of paths $X(\cdot)$
starting from position $y$ in space at time $0$ to arrive 
at position $x$ in space at time $t$,
{\sl formally} to be given by the infinite product of 
continously-many number of the Lebesgue measures $dX(\tau)$ 
on space ${\bf R}^d$ for each individual $\tau$ : 
$${\cal D}[X] := 
``{\rm constant}" \times \prod_{0 < \tau < t} dX(\tau). 
$$
Here the ``constant" should be something like 
$\big(\frac{i m}{2\pi \hbar(\delta t)}\big)
^{\frac{d}{2}\cdot(\frac{t}{\delta t}-1)}$
with $\delta t$ being some infinitesimal quantity of time, 
if one dares to try to write it, wondering what it means at all, 
but one may infer something from around (4.37) below.   
The right-hand side of (4.33) is what is called {\sl Feynman path integral} 
or, nowadays simply, {\sl path integral}.

To explain this, Feynman put the following {\sl Two Postulates} which turn out
to be equivalent to get the above expression (4.33) for $K(x,t; y,0)$, 
so that, for $f,\, g \in L^2({\bf R}^d)$, 
$$(f, \psi(\cdot,t)) 
= (f, e^{-\frac{it}{\hbar}[-\frac{\hbar^2}{2m} \Delta + V]}g)
= \int\int \overline{f(x)} K(x,t;y,0) g(y) dxdy.
$$

\noindent
{\sl Feynman's Two Postulates}

(i) $K(x,t; y,0)$ is the {\sl total probability amplitude} for the event
that the particle starts from position $y$ at time $0$ and arrives 
at position $x$ at time $t$. 
If $\varphi [X]$ stands for the {\sl probabilty amplitude} for the event
that it does this motion {\sl along each individual path} $X(\cdot)$,
$K(x,t; y,0)$ is the sum of the $\varphi [X]$ over all 
these paths $X(\cdot)$ :
\begin{equation}
 K(x,t; y,0) = \sum_{\{X: \, X(0) = y, X(t) = x\}} \varphi [X]. 
\end{equation}

(ii) The contribution $\varphi[X]$ from each $X(\cdot)$ to the total 
probabilty amplitude $K(x,t; y,0)$ is given by 
\begin{equation}  \varphi [X] = C e^{\frac{i}{\hbar}S(X)},
\end{equation}
where $C$ is a constant independent of path $X$.

These two postulates can be paraphrased:  In quantum mechanics there 
rules the Principle of Democracy that each individual path $X(\cdot)$ 
contributes to the total probabilty amplitude $K(x,t; y,0)$ with 
{\sl equal} {\sl weight} ({\it absolute value} in mathematics) and 
its personality is expressed by its {\sl phase} ({\it argument}  
in mathematics).

In this respect, in classical mechanics there does not rule 
Principle of Democracy, 
because the particle takes a {\sl particular path} between two space-time 
points $(y,0)$ and $(x,t)$ which makes the action $S(X)$ stationary,
called {\sl classical trajectory}. It is the path determined by 
Euler--Lagrange equation or, in the present case, Newton's equation of motion
: $m\frac{d^2}{ds^2} X (s) = -\nabla V(X(s))$.

The most important characteristic feature of these postulates lies in 
equation (4.36), which says that the amplitude $\varphi[X]$ is 
propotional to  the phase $e^{\frac{i}{\hbar}S(X)}$. The phrase  
``propotional to"  is that which Feynman determined to substitute 
for what Dirac had meant by the phrase ``analogous to" in 
[Di1-33, 2-35], [Di3-45] 
far before Feynman, by showing after his own analysis and deliberation that 
{\sl indeed this exponential function could be used in this manner directly} 
(see Preface of [FeyHi-65]). 

In classical mechanical circumstances, $\hbar$ is so small compared 
with other physical quantities that one may ignore and think of it as zero. 
The amazing thing is that this `integral' (4.33) can let us see how 
the transition is going to classical mechanics as $\hbar$ tends to zero. 
Namely, when $\hbar$ tends to zero, if the stationary
phase method should be valid for this `integral' (4.33), then the 
`integral' would turn out to receive the most crucial contribution 
from the path which makes the action $S(X)$ stationary, 
i.e. the classical trajectory (mentioned above)   
and its neighboring paths.  

\medskip\noindent
4.2.2. How to make it mathematics ?

Here we refer, among others,  only to two methods; one is  
by finite-dimensional approximation, and the other by 
imaginary-time path integral. In fact, it is by the first method that  
Feynman himself confirmed his idea of path integral. He calculated  
$K(x,t; y,0)$ by time-sliced approximation, 
making partition of the time interval $[0,t]$:   
$0=t_0 < t_1 < \cdots < t_n=t, \, (t_j-t_{j-1}=t/n)$, 
$x_j:=X(t_j)$, $\, x_0=X(0)=y,\, x_n=X(t)=x$, 
as the limit of
\begin{equation}
K_n(x,t;y,0) := \frac{\int_{({\bf R^d})^{n-1}}\!\!\exp\{\frac{it}{\hbar n}
\!\sum_{j=0}^{n-1}
[\frac{m}2(\frac{x_{j+1}-x_j}{{t}/{n}})^2-V(x_j)]\}
                          dx_1\cdot\cdots\cdot dx_{n-1}}
{\int_{({\bf R^d})^{n}}\!\!\exp\{\frac{it}{\hbar n}
\!\sum_{j=0}^{n-1}
 \frac{m}2(\frac{x_{j+1}-x_j}{t/n})^2\}dx_1\cdot\cdots\cdot dx_{n-1}dx_n} 
\end{equation}
as $n \rightarrow \infty$, 
to ascertain it to satisfy the Schr\"odinger equation (4.32).
Note that the denominator of the right-hand side of (4.37) is equal to
$\big(\frac{2\pi i\hbar \frac{t}{n}}{m}\big)^{\frac{d}{2}\cdot n}$.

\medskip
The second method is the one which the present article is mainly 
concerning.
We note with (4.33) that the solution $\psi(x,t)$ of the Schr\"odinger 
equation (4.32) turns out to be given by a heuristic path integral 
\begin{equation}
\psi(x,t)
= \int_{{\bf R}^d} K(x,t; y,0) f(y) dy
= \int_{\{X: \, X(t) = x \}} e^{\frac{i}{\hbar} S(X)}f(X(0)) {\cal D}[X].
\end{equation}
However, one should know that the `measure' 
${\cal D}[X]$ itself in general does not exist in this situation 
as a countably additive measure. Therefore we cannot go further. 
But if we rotate everthing  by $- 90^o$ : 
$t \rightarrow -it$ (real-time $t$ to imaginary-time $-it$) 
in complex $t$-plane (see Figure 1), 
i.e. if we go from our {\sl Minkowski} space-time 
to {\sl Euclidian} space-time, the situation will change.
Before actually doing it, for simplify put $\hbar =1$.  
Then, as our rotation also converts $ds$ to $-ids$, so does it  
$\dot{X}(s) = \frac{dX(s)}{ds}$ to $i\dot{X}(s) =\frac{dX(s)}{-ids}$ 
[where we don't mind thinking of ``$X(-is)$" as $X(s)$ again], 
so that $iS(X)$, the action $S(X)$ in (4.3) mutiplied by $i=\sqrt{-1}$,
is converted to time integral of the {\sl Hamiltonian} : 
$-\!\int_0^t[\frac{m}{2}\dot{X}(s)^2\!+\!V(X(s))]ds$.
Simultaneously, our (real-time) Schr\"odinger equation (4.32) 
is converted to the imaginary-time Schr\"odinger equation, i.e. 
heat equation [where writing $u(x,t)$ for $\psi(x,-it)$]:
\begin{equation}
 \frac{\partial}{\partial t} u(x,t)
 = \big[\frac{1}{2m}\Delta -V(x)\big] u(x,t),
\quad t>0, \quad x \in {\bf R}^d.
\end{equation}
Now we are going to get to the so-called {\sl Feynman--Kac formula}.
To this end, we replace the paths used so far by the time-reversed paths 
$X_0(s) := X(t-s), \,\, 0\leq s \leq t$, so that
 $X_0(0) =X(t)=x,\,\, X_0(y)= X(0)=y$.
Then $K(x,t;y,0)$ is changed to
\begin{equation}
K^E(x,t;y,0) := \int_{\{X_0:\, X_0(0) =x, X_0(t) = y\}}
 e^{-\int_0^t[\frac{m}{2}\dot{X}_0(s)^2+V(X_0(s))]ds}
{\cal D}[X_0].
\end{equation}
where the superscript ``$E$" is attributed to ``Euclidian", and
$K^E(x,t;y,0)$ should become the integral kernel for the heat 
equation (4.39).  
In passing we quickly insert here: if one were to follow the first method 
by Feynman as (4.37), one could also define $K^E(x,t;y,0)$ as the limit as 
$n\rightarrow \infty$ of
\begin{equation}
K^E_n(x,t;y,0) := \frac{\int_{({\bf R^d})^{n-1}}\!\!\exp\{-\frac{t}{n}
\!\sum_{j=0}^{n-1}
[\frac{m}2(\frac{x_{j+1}-x_j}{{t}/{n}})^2+V(x_j)]\}
                          dx_1\cdot\cdots\cdot dx_{n-1}}
{\int_{({\bf R^d})^{n}}\!\!\exp\{-\frac{t}{n}
\!\sum_{j=0}^{n-1}
 \frac{m}2(\frac{x_{j+1}-x_j}{t/n})^2\}dx_1\cdot\cdots\cdot dx_{n-1}dx_n}\,, 
\end{equation}
where  $ t_j-t_{j-1} = t/n\, (j=1,2, \dots,n);\,
x_j =X_0(t_j),\, x_0=X_0(0)=x,\, x_n=X_0(t)=y$.

We infer from (4.40) that
 the solution of the Cauchy problem for (4.39) with initial data 
$u(x,0)=g(x)$ should be given by the following path integral
\begin{eqnarray}
u(x,t) 
&=& \int_{{\bf R}^d} K^E(x,t;y,0)g(y)dy \nonumber\\
&=& \int_{\{X_0:\, X_0(0)=x\}}
e^{-\int_0^t[\frac{m}{2}\dot{X}_0(s)^2+V(X_0(s))]ds}g(X_0(t))
{\cal D}[X_0].
\end{eqnarray}
Here we have tacitly identified the two `integrals':
$$
\int_{{\bf R}^d} dy \int_{\{X_0:\, X_0(0) =x, X_0(t) = y\}} \cdots {\cal D}[X_0]
\quad \sim \quad
\int_{\{X_0:\, X_0(0)=x\}} \cdots {\cal D}[X_0].
$$
Needless to say, when the scalar potential 
$V(x)$ is absent, $K^E(x,t;y,0)$ becomes the heat kernel
$\big(\frac{2\pi t}{m}\big)^{-d/2} e^{-\frac{m}{2t}|x-y|^2}$, 
which is obtained as the inverse Fourier transform of the left-hand side of (4.29), 
or by calculating the integrals (4.41) and taking the limit $n \rightarrow \infty$. 
Note that the denominator of (4.41) is equal to 
$\big(\frac{2\pi \frac{t}{n}}{m}\big)^{\frac{d}{2}\cdot n}$.
By (4.40) we also see it have the following heuristic expression on the right-hand side
$$
 \frac{e^{-\frac{m}{2t}|x-y|^2}}{\big(\frac{2\pi t}{m}\big)^{d/2}}
= \int_{\{X_0:\, X_0(0) =x, X_0(t) = y\}}
 e^{-\int_0^t \frac{m}{2}\dot{X}_0(s)^2ds}{\cal D}[X_0].
$$

Remarkable is that Wiener, already around 1923, had constructed, 
for each individual $x \in {\bf R}^d$, a countably additive measure $\mu_x$ 
with $m=1$ (but of course valid for every $m>0$) 
 on the space $C_x :=C_x([0,\infty)\!\rightarrow \!{\bf R}^d)$
of the  continuous paths ({\sl Brownian motions}) 
$B: [0,\infty) \ni s \mapsto B(s) \in {\bf R}^d$ 
starting from $B(0)\!=\!x$ at time $t =0$. Further 
$\mu_x$ is a probability measure on $C_x$ with characteristic 
function (4.29), and now is called {\it Wiener measure}.

Around 1947, Kac, who had been at Cornell University as Feynman and 
heard his lecture at the Physics Colloquium, struck upon the very idea
of using the Wiener measure to represent the solution $u(x,t)$ of 
the Cauchy problem for the heat equation (4.39) (with $m=1$) with initial data 
$u(x,0)=g(x)$ as a first mathematical rigorous, 
{\sl genuine functional integral}
\begin{equation}
 u(x,t)= \int K^E(x,t;y,0)g(y)dy
 = \int_{C_x([0,\infty)\rightarrow{\bf R}^d)}
         \!\!\!e^{-\int_0^t V(B(s)) ds} g(B(t))d\mu_x(B),
\end{equation}
the same formula as (4.1) already mentioned at the top of this section.
This is the {\sl Feynman--Kac formula} [Kac-66/80] mentioned in advance. 
Thus, identify the path $X_0(\cdot)$ appearing on the 
right-hand side of 
(4.40)/(4.42) with the continuous path $B(\cdot)$ in the space 
$C_x([0,\infty)\!\rightarrow\!{\bf R}^d)$, then 
the Wiener measure $\mu_x$ turns out to be constructed from 
the factor 
``$\,e^{-\int_0^t\frac{m}{2}\dot{B}(s)^2ds}\, {\cal D}[B]\,$" 
(with $m=1$) on the right-hand side of (4.40)/(4.42).

\medskip\noindent
4.2.3. The case for relavistic Schr\"odinger equation.

We begin with the relativistic Schr\"odinger equation for a relativistic 
particle of mass $m$ with positive energy in an electromagnetic field: 
\begin{equation}
i \hbar \frac{\partial}{\partial t} \psi (x,t)= [H-m] \psi(x,t), 
   \quad t > 0, \quad x  \in {\bf R}^d,
\end{equation}
where $H$ is one of the relativistic Schr\"odinger operators 
$H^{(1)},\, H^{(2)}$ and $H^{(3)}$ 
corresponding to the classical symbol $\sqrt{(\xi-A(x))^2+m^2} + V(x)$. 
This equation (4.44) was already briefly mentioned at the top of Section 4.1.
However, here for the moment we go with the quantization 
``$\xi \rightarrow -i\hbar\nabla$" with $\hbar$ recovered, 
but not ``$\xi \rightarrow -i\nabla$" used there.
In this case it is more appropriate to use the method of 
{\sl phase space path integral} or {\sl Hamiltonian path integral}
 (Feynman [Fey-65, p.125], Garrod [G-66], Mizrahi [M-78]):
\begin{equation}
\displaystyle{\int} e^{\frac{i}{\hbar}S(\Xi,X)}{\cal D}[\Xi]{\cal D}[X]
\end{equation}
with a `measure' ${\cal D}[\Xi]{\cal D}[X]$ on the space of phase space 
paths $(\Xi, X)$, pairs of momentum path $\Xi(s)$ and
position path $X(s)$ on the phase space ${\bf R}^d \times {\bf R}^d$, and
with {\sl action} written with each pair $(\Xi(s), X(s))$.
Then, under this circumstance, the previous path integral (4.31), (4.33) 
in the nonrelativistic case is also called {\sl configuration path integral}. 

The solution $\psi(x,t)$ of the Cauchy problem for (4.44)
with initial data $\psi(x,0) = f(x)$
can be written as $\psi(x,t) = \int K(x,t;y,0) f(y) dy$ with
integral kernel $K(x,t;y,0)$ called {\sl fundamental solution or propagator}. 
Then the method of phase space path integral or Hamiltonian path integral 
assumes $K(x,t;y,0)$ to have the following path integral representaion:
\begin{equation}
K(x,t;y,0) = \int_{\{(\Xi,X);\, X(0)=y, X(t)=x,\, \Xi:\, {\rm arbitrary}\}} 
            e^{\frac{i}{\hbar}S(\Xi(s),X(s))}{\cal D}[\Xi]{\cal D}[X]\,.
\end{equation}
Here the {\sl action} $S(\Xi,X)$ along the phase space path $(\Xi,X)$ is given by
\begin{equation}
S(\Xi,X) = \displaystyle{\int}_0^t \Big[\Xi(s)\cdot\dot{X}(s) 
          -\Big(\sqrt{(\Xi(s)-A(X(s)))^2+m^2} -m + V(X(s))\Big)\Big]ds,
\end{equation}
where $\dot{X}(s) = \frac{d}{ds}X(s)$ in the same way 
as in the nonrelativistic case (4.34). 
${\cal D}[\Xi]{\cal D}[X]$ is a uniform `measure', if it exists,
on the space of phase space paths 
$(\Xi, X): [0,t] \ni s \mapsto (\Xi(s), X(s)) \in {\bf R}^d \times {\bf R}^d$
with $X(0)=y, X(t)=x$, but 
$\Xi$ being {\rm unrestricted\,\, and\,\, so\,\, arbitrary}, 
{\sl formally} to be given by the infinite product of continously-many number 
of the Lebesgue measures $d\Xi(\tau)dX(\tau)\,$ 
(precisely, divided by $(2\pi)^d$) on phase space
${\bf R}^{2d} = {\bf R}^d \times {\bf R}^d$ for each individual $\tau$ :
$$
{\cal D}[\Xi]{\cal D}[X] := 
\prod_{0< \tau < t}\frac{d\Xi(\tau)dX(\tau)}{(2\pi)^{d}}.
$$ 
In this `integral' (4.46) we make the transform of variables (paths):  
$\Xi'(s) = \Xi(s) -A(X(s))$ and $X'(s) = X(s)$, where we note the 
{\sl formal} Jacobi determinant 
$\frac{\partial(\Xi(s)\,, \,\,X(s))}{\partial(\Xi'(s),X'(s))}$ of 
this transform is $1$. 
Write $\Xi(s)$ for $\Xi'(s)$ and $X(s)$ for $X'(s)$ again, 
then (4.46) becomes 
\begin{eqnarray}
\!\!\!\!\!\!K(x,t;y,0) \!\!\!\!&\!=\!&\!\!\!\!\displaystyle
{\int}_{\!\!\{(\Xi,X);\, X(0)=y, X(t)=x,\, \Xi:\, {\rm arbitrary}\}} 
  \nonumber\\ 
\!\!\!\!&\!\!&\!\!\!\!
\times e^{\frac{i}{\hbar}\int_0^t\big[(\Xi(s)+A(X(s))\cdot\dot{X}(s)
        -\big(\sqrt{\Xi(s)^2+m^2}-m + V(X(s))\big)\big]ds} 
                                      {\cal D}[\Xi]{\cal D}[X]. 
\end{eqnarray}
We want to find a path integral formula for the solution of 
the imaginary-time magenetic relativistic Schr\"odinger equation (4.3). 
For simplicity we put 
$\hbar =1$ as before. We go from real time $t$ to imaginary time $-it$. 
This procedure also converts $ds$ to $ -ids$ and so 
$\dot{X}(s)= \frac{d}{ds}X(s)$ to $i\dot{X}(s)= \frac{dX(s)}{-ids}$ 
[where we don't mind thinking of ``$\Xi(-is),\, X(-is)$" as $\Xi(s),\, X(s)$, 
respectively, again]. Then we replace the phase space paths used so far 
by the time-reversed phase space paths
: $X_0(s) := X(t-s),\,\, \Xi_0(s) := \Xi(t-s),\,\, 0\leq s \leq t$,
so that $X_0(0) =X(t)=x,\,\, X_0(y)= X(0)=y$.
As a result, (4.46) is changed to 
\begin{eqnarray}
\!\!\!&\!\!\!&\!\!\!\!\!\!K^E(x,t;y,0) \nonumber\\
&\!\!\!:=&\!\!\! \displaystyle
{\int}_{\!\!\{(\Xi_0,X_0);\, X_0(0)=x,X_0(t)=y,\, 
       \Xi:\, {\rm arbitrary}\}}  \nonumber\\ 
\!\!\!\!&\!\!&\!\!\!\!
\times e^{\int_0^t\big[i(\Xi_0(s)+A(X_0(s))\cdot\dot{X}_0(s)
        -\big(\sqrt{\Xi_0(s)^2+m^2}-m + V(X_0(s))\big)\big]ds} 
                                   {\cal D}[\Xi_0]{\cal D}[X_0]. 
\end{eqnarray}
At this final stage we rewrite $\Xi_0(s),\, X_0(s)$ as $\Xi(s),\, X(s)$ 
again. Thus  we have heuristically arrived, for the solution $u(x,t)$ of 
the Cauchy problem for (4.3) with initial data $u(x,0)=g(x)$,  
at the following path integral representation: 
\begin{eqnarray}
u(x,t) \!\!\!\!&\!=\!&\!\!\!\! (e^{-t[H-m]}g)(x)
= \int K^E(x,t;y,0) g(y) dy \nonumber\\
\!\!\!\!&\!=\!&\!\!\!\! \displaystyle{\int}_{\!\{\! X(0)=x\}}
 e^{\int_0^t\big[i(\Xi(s)+A(X(s)))\cdot\dot{X}(s)
        -\big(\sqrt{\Xi(s)^2+m^2}-m + V(X(s))\big)\big]ds} 
                              g(X(t)){\cal D}[\Xi]{\cal D}[X]
\nonumber\\
\!\!\!\!&\!=\!&\!\!\!\! \displaystyle{\int}_{\!\{\! X(0)=x\}}
e^{\int_0^t\big[iA(X(s))\cdot\dot{X}(s) - V(X(s))\big]ds}
          \nonumber\\
&&\qquad\quad \times 
  e^{\int_0^t\big[i\Xi(s)\cdot\dot{X}(s)
                 -\big(\sqrt{\Xi(s)^2+m^2}-m\big)\big]ds} 
 g(X(t)){\cal D}[\Xi]{\cal D}[X].\qquad
\end{eqnarray}
Now we ask whether our path integral formulas, 
(4.9)/(4.10) in Theorem 4.1, (4.20)/(4.21) in Theorem 4.3 and  
 (4.27)/(4.28) in Theorem 4.5, can be well  derived or at least 
well inferred from this formal expression of `integral' (4.50). 
First of all, if both the vector and scalar potentials 
$A(x)$ and $V(x)$ are absent, (4.50) is reduced to
\begin{eqnarray}
u(x,t) 
&=& (e^{-t[\sqrt{-\Delta +m^2}-m]}g)(x) = \int k_0(x-y,t)g(y) dy
              \nonumber\\
&=& \displaystyle
  {\int}_{\!\{\! X(0)=x\}}e^{\int_0^t\big[i\Xi(s)\cdot\dot{X}(s)
  -\big(\sqrt{\Xi(s)^2+m^2}-m\big)\big]ds} 
                               g(X(t)){\cal D}[\Xi]{\cal D}[X],
\end{eqnarray}
where $k_0(x-y,t)$ is the integral kernel of the semigroup
 $e^{-t[\sqrt{-\Delta +m^2}-m]}$ in (3.3), and, 
similarly to the nonrelativistic case, we have 
identified the two `integrals':
\begin{eqnarray*}
&&\int_{{\bf R}^d} dy 
\int_{\{(\Xi,X);\, X(0)=x,X(t)=y,\, \Xi:\, {\rm arbitrary}\}} 
\cdots {\cal D}[\Xi]{\cal D}[X]\\
&&\quad \sim \quad
\int_{\{(\Xi,X);\, X(0)=x,\, \Xi:\, {\rm arbitrary}\}} 
\cdots {\cal D}[\Xi]{\cal D}[X].
\end{eqnarray*}
Noting that the second and/or third member of (4.50)
is equal to 
$$
\int_{D_x([0,\infty)\rightarrow{\bf R}^d)} g(X(t)) d\lambda_x(X),
$$
we see that the factor 
\begin{equation}
\exp\Bigl\{\int_0^t\Big[i\Xi(s)\cdot\dot{X}(s)
 -\Big(\sqrt{\Xi(s)^2+m^2}-m\Big)\Big]ds\Bigr\}{\cal D}[\Xi]{\cal D}[X]
\end{equation}
turns out to be identified with the 
probability measure $\lambda_x$, (4.4) introduced in Section  4.1,  
connected with the L\'evy process concerned. 
Next, we shall see that, since there is no problem
for the factor $e^{-\int_0^t V(X(s))ds}$, 
the problem lies only in how to interpret and understand the factor  
$$
e^{i\int_0^t A(X(s))\cdot\dot{X}(s)ds}
= \prod_{j=1}^n e^{i\int_{t_{j-1}}^{t_j} A(X(s))\cdot\dot{X}(s)ds},
$$
when dividing the time interval $[0,t]$ into $n$ equal small subintervals 
$[t_0,t_1], \dots, [t_{n-1},t_n]$ with 
$t_j= jt/n,\, j=0, 1, 2, \dots, n$, in fact, whether,
for small interval $[t_{j-1}, t_j]$ or large $n$, the factor 
$e^{i\int_{t_{j-1}}^{t_j} A(X(s))\cdot\dot{X}(s)ds}$ can allow a good
 approximation to be suggested by the obtained path integral formulas
for $H^{(1)}$, $H^{(2)}$ and $H^{(3)}$.

\medskip\noindent
(1) First we consider the case for $H^{(1)}$ by  approximating
the factor 
$$
\exp\Big[i\int_{t_{j-1}}^{t_j} A(X(s))\cdot\dot{X}(s)ds\Big]
\quad \hbox{\rm by}\quad 
\exp\Big[iA\Big(\frac{X(t_{j-1})+X(t_j)}{2}\Big)
      \cdot (X(t_{j})-X(t_{j-1}))\Big]
$$
on each subinterval $[t_{j-1},t_j]$ (``midpoint prescription"). 
Then the last member of (4.15) is expected to be the limit 
$n\rightarrow \infty$ of 
\begin{eqnarray}
&&\displaystyle{\overbrace{\int_{{\bf R}^{2d}}\cdots\int_{{\bf R}^{2d}}}
^{\mbox{$n$ times}}}
\Big(e^{i\sum_{j=1}^n \,(\Xi(t_{j})
+A(\frac{X(t_{j-1})+X(t_j)}{2})\cdot(X(t_{j})-X(t_{j-1}))}
 \nonumber\\
&&\times e^{-\sum_{j=1}^n[\sqrt{\xi_{j}^2+m^2}-m 
+V(\frac{X(t_{j-1})+X(t_j)}{2})]\frac{t}{n}}\Big) g(X(t_n)) 
 \prod_{j=1}^n \frac{d\Xi(t_{j})dX(t_{j})}{(2\pi)^{d}} \nonumber\\ 
&=&\displaystyle{\overbrace{\int_{{\bf R}^{2d}}\cdots\int_{{\bf R}^{2d}}}
^{\mbox{$n$ times}}}
\Big(e^{i\sum_{j=1}^n \,[(\Xi(t_{j})\cdot(X(t_{j})-X(t_{j-1}))
 -(\sqrt{\xi_{j}^2+m^2}-m)\frac{t}{n}]}
 \nonumber\\
&&\qquad\qquad \times
 e^{i\sum_{l=1}^n[A(\frac{X(t_{l-1})+X(t_l)}{2})\cdot(X(t_{j})-X(t_{j-1}))
   -V(\frac{X(t_{l-1})+X(t_l)}{2})\frac{t}{n}]}\Big) \nonumber \\
 &&\qquad\qquad 
  \times g(X(t_n)) 
  \prod_{j=1}^n \frac{d\Xi(t_{j-1})dX(t_{j-1})}{(2\pi)^{d}},
\qquad X(0)=X(t_0) =x. \qquad
\end{eqnarray}
Then putting $\xi_j=\Xi(t_j),\ x_j=X(t_j)$ makes (4.50) equal to
\begin{eqnarray}
&&\displaystyle{\overbrace{\int_{{\bf R}^{2d}}\cdots\int_{{\bf R}^{2d}}}
^{\mbox{$n$ times}}}
\prod_{j=1}^n e^{i(x_j-x_{j-1})\cdot\xi_{j}}
e^{-[\sqrt{\xi_{j}^2+m^2}-m]\frac{t}{k}} \nonumber\\
&&\qquad
\times \exp\Big\{i\sum_{l=1}^n 
   \Big[A\Big(\frac{x_{l-1}+x_l}{2})\cdot(x_l-x_{l-1}) 
           -V\Big(\frac{x_{l-1}+x_l}{2})\frac{t}{n}\Big)\Big]\Big\}
 \nonumber\\
&&\qquad \times g(x_n) \prod_{j=1}^n \frac{d\xi_{j}dx_{j}}{(2\pi)^d}, 
\qquad x_0 =x.
\end{eqnarray}
Performing all the $d\xi_j$ integrals yields
\begin{eqnarray}
&&\displaystyle{\overbrace{\int_{{\bf R}^{d}}\cdots\int_{{\bf R}^{d}}}
^{\mbox{$n$ times}}}
k_0(x_0-x_1,t/n)k_0(x_1-x_2,t/n)\cdot\cdots\cdot 
                       k_0(x_{n-1}-x_n,t/n)\nonumber\\
&&\times \exp\Big\{-\sum_{l=1}^n
\Big[iA\Big(\frac{x_{l-1}+x_l}{2}\Big)\cdot(x_{l-1}-x_{l}) 
  +V\big(\frac{x_{l-1}+x_l}{2}\big)\frac{t}{n}\Big)\Big]\nonumber\\
&&\times g(x_n) dx_1\cdot \cdots \cdot dx_{n}, \qquad x_0 =x,
\end{eqnarray}
where $k_0(x,t)$ is the integral kernel of $e^{-t[\sqrt{-\Delta+m^2}-m]}$
in (3.3). Note that (4.55) is the same as the second member of (4.15), 
which was shown in Proposition 4.2 
to converge to $e^{-t[H^{(1)}-m]}g$. Therefore we may think 
that the expression (4.9) with (4.10) is heuristically connected with 
(4.50) in the limit $n\rightarrow \infty$ of the expression (4.55) 
which should be the path integral formula for $e^{-t[H^{(1)}-m]}g$
in Theorem 4.1.

\medskip\noindent
(2) Next we consider the case for $H^{(2)}$ by approximating the factor 
$$
\exp\Big[i\int_{t_{j-1}}^{t_j} A(X(s))\cdot\dot{X}(s)ds\Big]
$$
{\sl by} 
$$
\exp\Big[i\int_0^1 A\big((1-\theta)X(t_{j})+\theta X(t_{j-1})\big)
\cdot (X(t_{j})-X(t_{j-1})) d\theta\Big]
$$
on each subinterval $[t_{j-1},t_j]$. 
The same arguments as in (1) above will show 
the expression (4.20) with (4.21) is also heuristicaly 
connected with (4.50), leading to the path integral formula 
(4.20) with (4.21) for $e^{-t[H^{(2)}-m]}g$ in Theorem 4.3.

\medskip\noindent
(3) Finally, we come to the case for $H^{(3)}$. Indeed, (4.27)/(4.28)
is a mathematically rigorous, beautiful path integral
but it does not seem to be one which can be heuritically deduced 
from the formal expression  of `integral' (4.50). 
We could not think the factor 
$\exp\Big[i\int_{t_{j-1}}^{t_j} A(X(s))\cdot\dot{X}(s)ds\Big]$
can allow a good approximation to be suggested by (4.27)/(4.28) 
for $H^{(3)}$. 
It is because $H_A^{(3)}$ does not seem to be so explicitly well
expressed by a pseudo-differential operator defined through a certain 
{\sl tractable symbol} as $H_A^{(1)}$ and $H_A^{(2)}$.

\section{Summary}

Finally, we will collect here, as summary, the three path integral 
representation 
formulas in Theorems 4.1, 4.3, 4.5 so as to be able to explicitly see 
how they are $x$-dependent. To do so, we replace the $x$-dependent path space 
$D_x \equiv D_x([0,\infty)\rightarrow {\bf R}^d)$ / 
$C_x \equiv C_x([0,\infty)\rightarrow {\bf R}^d)$ 
(with probability measure $\lambda_x$ / $\mu_x$) 
{\sl by} the $x$-independent path space 
$D_0 \equiv D_0([0,\infty)\rightarrow {\bf R}^d)$ / 
$C_0 \equiv C_0([0,\infty)\rightarrow {\bf R}^d)$ of the paths $X(s)$ / $B(s)$
starting from $0$ in space ${\bf R}^d$ at time $s=0$ 
(with probability measure $\lambda_0$ / $\mu_0$), respectively.
Namely, in the path integral representation formulas in these three theorems, 
we make change of space, probablity measure and paths by translation $x$: 

\smallskip
$D_x \!\!\rightarrow\!\! D_0$, $\lambda_x\!\!\rightarrow\!\! \lambda_0$,
$X(s)\!\! \rightarrow\!\! X(s)\!+\!x$, 

$C_x \!\!\rightarrow\!\! C_0$, $\mu_x\!\!\rightarrow\!\! 
\mu_0$,
$B(s)\!\! \rightarrow\!\! B(s)\!+\!x$,
$B(T(s))\!\! \rightarrow\!\! B(T(s))\!+\!x$,

\medskip
{\footnotesize
\begin{eqnarray*}
\hbox{(4.9)}&:& (e^{-t[H^{(1)}-m]}g)(x) = 
 \int_{D_0([0,\infty)\rightarrow {\bf R}^d)} e^{-S^{(1)}(X;x,t)} 
  g(X(t)+x)\,d\lambda_0 (X),\\ 
&&S^{(1)}(X;x,t)= i\int_0^{t+}\int_{|y|>0} 
        A(X(s-)+x+\frac{y}{2})\!\cdot\! y\, \widetilde{N}_X(dsdy)\nonumber\\
&&\qquad\qquad\qquad\quad +i\int_0^{t} \int_{|y|>0} 
        [A(X(s)+x+\frac{y}{2})-A(X(s)+x)]\!\cdot\! y\, dsn(dy) \nonumber\\
&&\qquad\qquad\qquad\quad    +\int_0^t V(X(s)+x)ds\,;\\
\hbox{(4.20)}&:& (e^{-t[H^{(2)}-m]}g)(x) = 
  \int_{D_0([0,\infty)\rightarrow {\bf R}^d)} e^{-S^{(2)}(X;x,t)} 
        g(X(t)+x)\,d\lambda_0 (X),\\ 
&&S^{(2)}(X;x,t) = i\int_0^{t+}\int_{|y|>0} 
  \Big(\int_0^1 A(X(s-)\!+\!x\!+\!\theta y)d\theta\Big) \!\cdot\! y\, 
               \widetilde{N}_X(dsdy)\nonumber\\
 &&\qquad\qquad\qquad\quad
   +i\int_0^{t}\!\int_{|y|>0}
 \!\!\Big[\!\int_0^1\!\! A(X(s\!)\!+\!x\!+\!\theta y)d\theta
           -A(X(s)\Big]\!\cdot\!y \,dsn(dy) \nonumber\\
&&\qquad\qquad\qquad\quad
        \!+\!\!\int_0^t \!\!V(X(s)\!+\!x)ds\,; \nonumber\\ 
\hbox{(4.27)}&:& (e^{-t[H^{(3)]}-m]}g)(x) =
 \int\int_{\stackrel{\scriptstyle C_0([0,\infty)\rightarrow {\bf R}^d)}
{\times D_0([0,\infty)\rightarrow{\bf R})}} \,e^{-S^{(3)}(B,T;x,t)}
    g(B(T(t))\!+\!x) \,d\mu_0(B) d\nu_0(T),\\ 
&&S^{(3)}(B,T;x,t) = 
 i\int_0^{T(t)} \!\!\!\!\!A(B(s)\!+\!x)\!\cdot \!dB(s)
   \!+\!\frac{i}{2}\!\int_0^{T(t)}\!\!\!\!\! \hbox{\rm div} A(B(s)\!+\!x)ds
                                                                \nonumber\\ 
 &&\qquad\qquad\qquad\qquad   +\!\! \int_0^t \!\!V(B(T(s))\!+\!x)ds,\nonumber\\
 &&\qquad\qquad\qquad \equiv 
  i\int_0^{T(t)} A(B(s)+x)\circ dB(s)+ \int_0^tV(B(T(s))+x)ds 
\end{eqnarray*}
}

\section{Acknowledgments}
It is a great pleasure for me to contribute to this volume in the series
``Advances in Partial Differential Equations". 
My sincere thanks go to Yuji Kasahara for enlightening and helpful 
discussions about the Remark after Theorem 4.5.
I am very grateful to the Editors Michael Demuth and Werner Kirsch, 
not only for their kind invitation but also for their generous and warm patience 
to wait for the arrival of my delayed manuscript. 
This work is supported in part by JSPS Grant-in-Aid for Scientific Research 
No. 23540191.



\end{document}